\documentclass[a4paper]{aa}
\usepackage[T1]{fontenc}
\usepackage[utf8]{inputenc}
\usepackage{graphicx}
\usepackage{color}
\usepackage{txfonts}
\usepackage{microtype}
\usepackage{ellipsis}
\usepackage{natbib}

\bibpunct{(}{)}{;}{a}{}{,}
\usepackage[pdftex,colorlinks=true,urlcolor=blue]{hyperref}
\usepackage[xindy,nonumberlist,nopostdot,nogroupskip]{glossaries}

\makeglossaries
\usepackage[xindy]{imakeidx}
\makeindex

\begin{document} 
\title{How the Vikings Navigated With the Sun}
\author{M. Nielbock}
\institute{Haus der Astronomie, Campus MPIA, Königstuhl 17, D-69117 
Heidelberg, Germany\\
\email{nielbock@hda-hd.de}}

\date{Received July 14, 2016; accepted }

\abstract{The students are introduced to navigation in general, and, in particular, the skills the medieval people of the Vikings used to navigate on the open sea. The topics of navigation and the Vikings are introduced by questions and a short story. A hands-on activity illustrates to the students how the navigational tool of the shadow board might have helped the Vikings to determine the cardinal directions and to sail along latitude. A miniature version of a shadow board simulates how the shadow cast by the Sun was probably used for this. Finally, a simple math activity for more experience students demonstrates the geometry that is involved in that technique.}

\keywords{Earth, navigation, astronomy,  history, geography, Sun, equator, latitude, longitude, meridian,  celestial navigation, Vikings, seasons, sundial}

\maketitle
%

\section{Background information}
\subsection{Latitude and longitude}
\begin{figure}[!ht]
 \resizebox{\hsize}{!}{\includegraphics{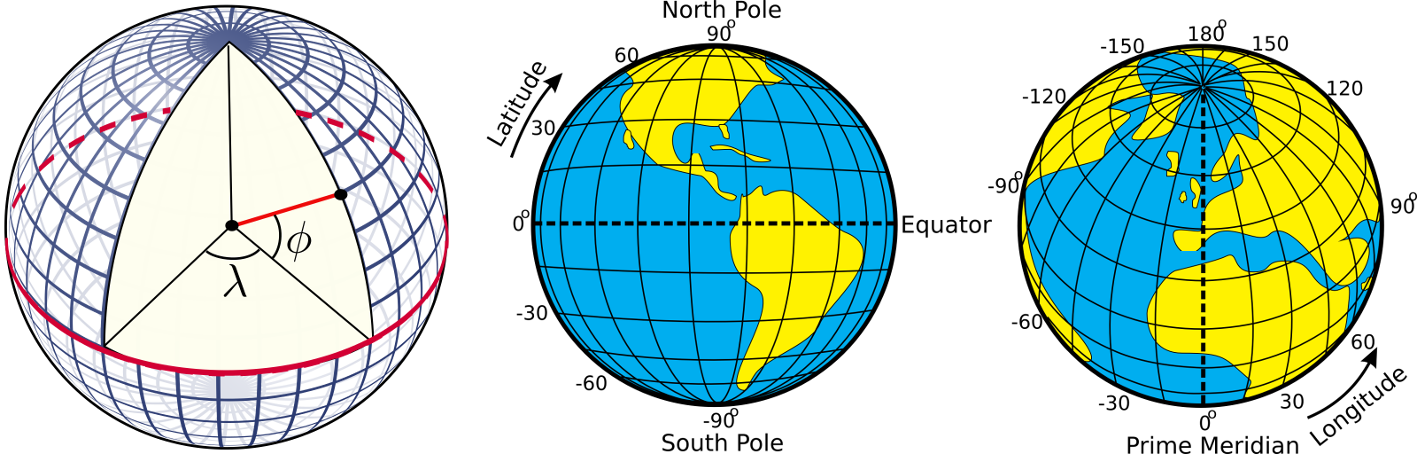}}
 \caption{Illustration of how the latitudes and longitudes of the Earth are defined (Peter Mercator, djexplo, CC0).}
 \label{f:latlong}
\end{figure}

Any location on an area is defined by two coordinates. The surface of a sphere is a curved area, but using coordinates like up and down does not make much sense, because the surface of a sphere has neither a beginning nor an ending. Instead, we can use
\newglossaryentry{spherical}
{
         name = {Spherical polar coordinates},
  description = {The natural coordinate system of a flat plane is Cartesian and       measures distances in two perpendicular directions (ahead, back, left, right). For a sphere, this is not very useful, because it has neither beginning nor ending. Instead, the fixed point is the centre of the sphere. When projected outside from the central position, any point on the surface of the sphere can be determined by two angles with one of them being related to the symmetry axis. Such axis defines two poles. In addition, there is the radius that represents the third dimension of space, which permits determining each point within a sphere. This defines the spherical polar coordinates. When defining points on the surface of a sphere, the radius stays constant.}
}
spherical polar coordinates originating from the centre of the sphere with the radius being fixed (Fig.~\ref{f:latlong}). Two angular coordinates remain. Applied to the Earth, they are called the latitude and the longitude. Its rotation provides the symmetry axis. The North Pole is defined as the point, where the theoretical axis of rotation meets the surface of the sphere and the rotation is counter-clockwise when looking at the North Pole from above. The opposite point is the South Pole. The equator is defined as the great circle
\newglossaryentry{great}
{
         name = {Great circle},
  description = {A circle on a sphere, whose radius is identical to the radius of the sphere.}
}
half way between the two poles.

The latitudes are circles parallel to the equator. They are counted from $0\degr$ at the equator to $\pm 90\degr$ at the poles. The longitudes are great circles connecting the two poles of the Earth. For a given position on Earth, the longitude going through the
\newglossaryentry{zenit}
{
         name = {Zenith},
  description = {Point in the sky directly above.}
}
zenith, the point directly above, is called the meridian. This is the line the Sun apparently
\newglossaryentry{appa}
{
         name = {Apparent movement},
  description = {Movement of celestial objects in the sky which in fact is caused by the rotation of the Earth.}
}
crosses at local noon. The origin of this coordinate is defined as the
\newglossaryentry{meri}
{
         name = {Meridian},
  description = {A line that connects North and South at the horizon via the zenith.}
}
Prime Meridian, and passes Greenwich, where the Royal Observatory of England is located. From there, longitudes are counted from $0\degr$ to $+180\degr$ (eastward) and $-180\degr$ (westward). 

Example: Heidelberg in Germany is located at 49\fdg4 North and 8\fdg7 East.

\subsection{Elevation of the pole (pole height)}
\begin{figure}[!ht]
 \resizebox{\hsize}{!}{\includegraphics{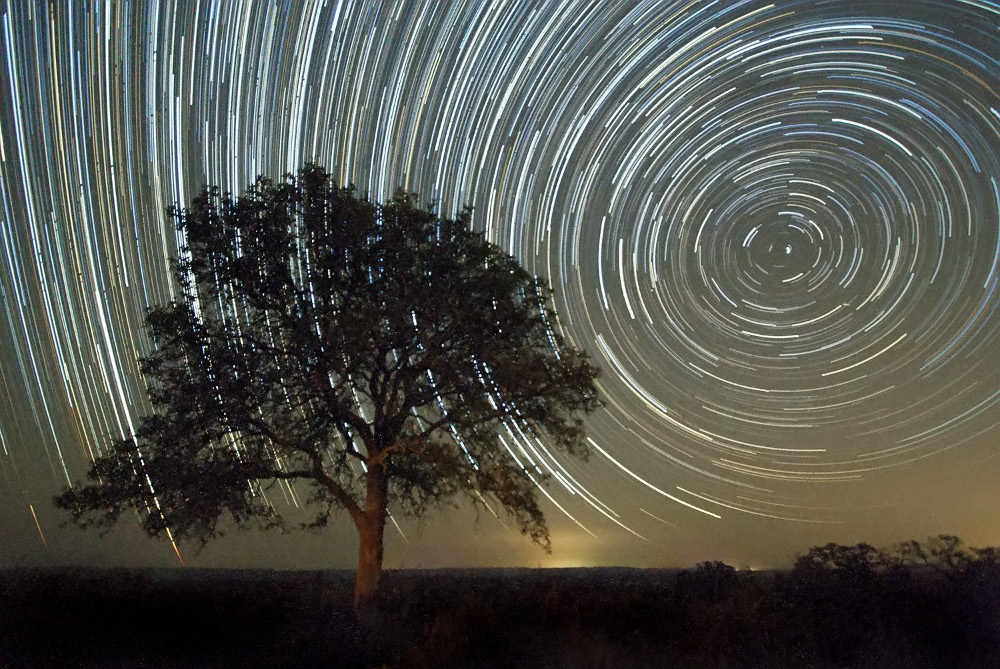}}
 \caption{Trails of stars at the sky after an exposure time of approximately 2 hours (Ralph Arvesen, Live Oak star trails, 
\url{https://www.flickr.com/photos/rarvesen/9494908143}, \url{https://creativecommons.org/licenses/by/2.0/legalcode}).}
 \label{f:trails}
\end{figure}

If we project the terrestrial coordinate system of latitudes and longitudes at the sky, we get the celestial equatorial coordinate system. The Earth's equator becomes the celestial equator and the geographic poles are extrapolated to build 
the celestial poles. If we were to make a photograph with a long exposure of the northern sky, we would see from the trails of the stars that they all revolve about a common point, the northern celestial pole (Fig.~\ref{f:trails}).

\begin{figure}[!ht]
 \centering
 \resizebox{0.45\hsize}{!}{\includegraphics{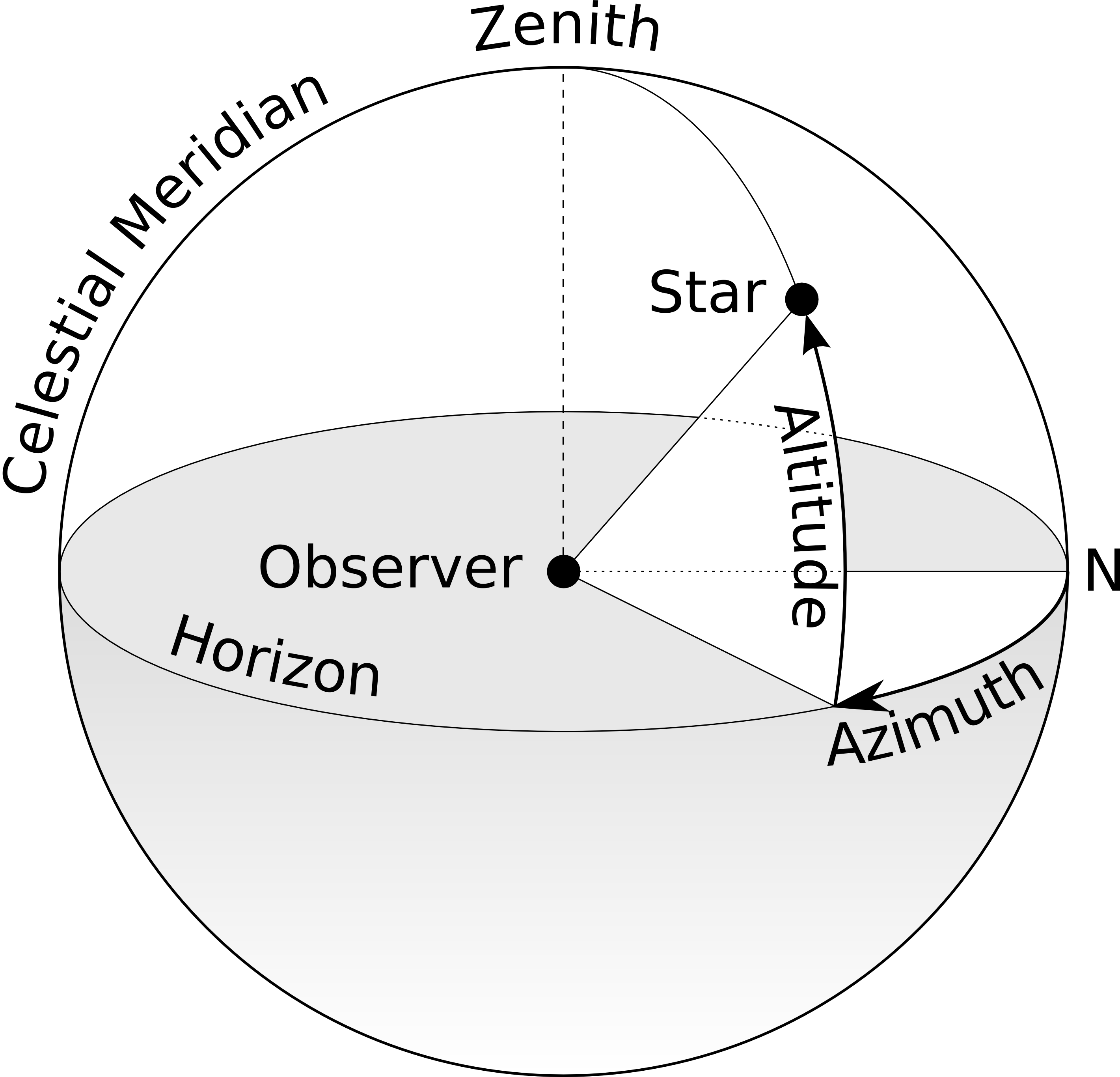}}
 \caption{Illustration of the horizontal coordinate system. The observer is the origin of the coordinates assigned as azimuth and altitude or elevation (TWCarlson, \url{https://commons.wikimedia.org/wiki/File:Azimuth-Altitude_schematic.svg},
``Azimuth-Altitude schematic'', \url{https://creativecommons.org/licenses/by-sa/3.0/legalcode}).}
  \label{f:altaz}
\end{figure}

In the northern hemisphere, there is a moderately bright star near the celestial pole, the North Star or Polaris. At the southern celestial pole, there is no such star that can be observed with the naked eye. Other procedures have to be applied to find it. If we stood exactly at the geographic North Pole, Polaris would always be directly overhead. We can say that its elevation
\newglossaryentry{elev}
{
         name = {Elevation},
  description = {Angular distance between a celestial object and the horizon.}
}
would be (almost) $90\degr$. This information already introduces the horizontal coordinate system (Fig.~\ref{f:altaz}).

It is the natural reference we use every day. We, the observers, are the origin of that coordinate system located on a flat plane whose edge is the horizon. The sky is imagined as a hemisphere above. The angle between an object in the sky and the horizon is the altitude or elevation. The direction within the plane is given as an angle between $0\degr$ and $360\degr$, the azimuth, which is usually counted clockwise from north. In navigation, this is also called the bearing. The meridian is the line that connects North and South at the horizon and passes the zenith.

\begin{figure}[!ht]
 \resizebox{\hsize}{!}{\includegraphics{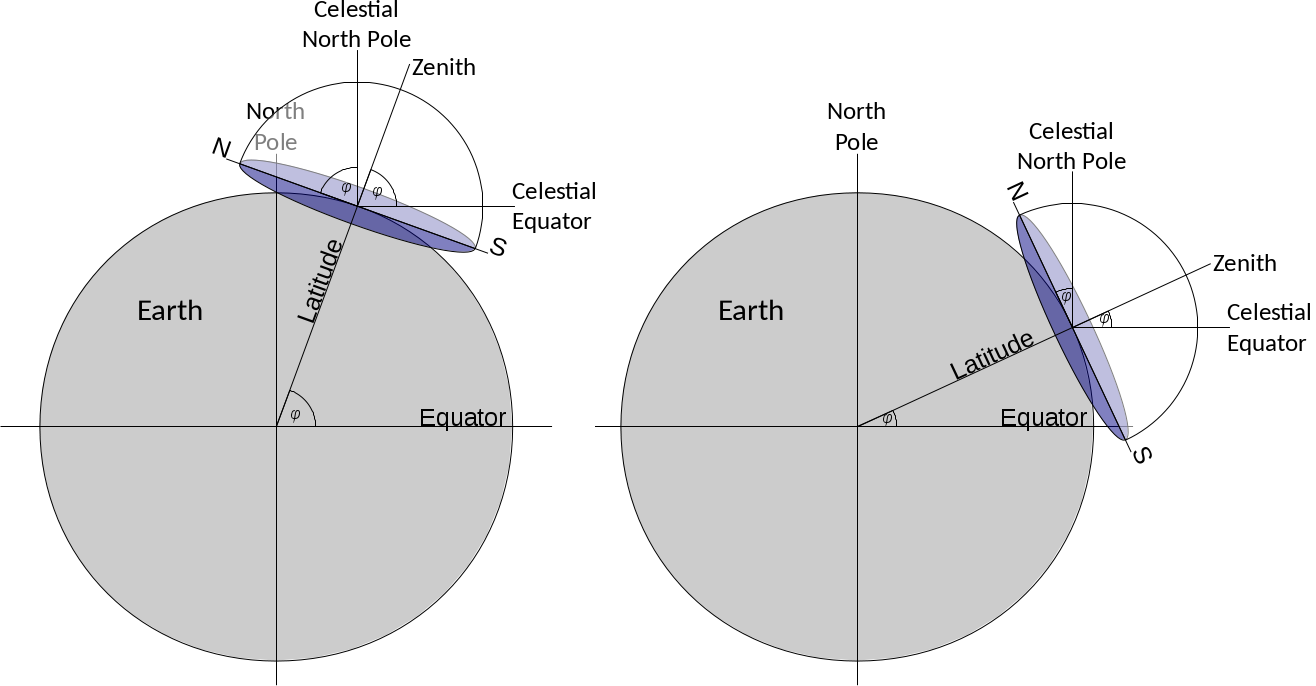}}
 \caption{When combining the three coordinate systems (terrestrial spherical, celestial equatorial, local horizontal), it becomes clear that the latitude of the observer is exactly the elevation of the celestial pole, also known as the pole height (own work).}
   \label{f:poleheight}
\end{figure}

For any other position on Earth, the celestial pole or Polaris would appear at an elevation smaller than $90\degr$. At the equator, it would just graze the horizon, i.e. be at an elevation of $0\degr$. The correlation between the latitude (North Pole = $90\degr$, Equator = $0\degr$) and the elevation of Polaris is no coincidence. Figure~\ref{f:poleheight} combines all three 
mentioned coordinate systems. For a given observer at any latitude on Earth, the local horizontal coordinate system touches the terrestrial spherical polar coordinate system at a single tangent point. The sketch demonstrates that the elevation of the celestial North Pole, also called the
\newglossaryentry{poleht}
{
         name = {Pole height},
  description = {Elevation of a celestial pole. Its value is identical to the latitude of the observer on Earth.}
}
pole height, is exactly the northern latitude of the observer on Earth.

\subsection{Early navigational skills}
Early seafaring peoples often navigated along coastlines before sophisticated navigational skills were developed and tools were invented. Sailing directions helped to identify coastal landmarks \citep{hertel_geheimnis_1990}. To some extent, their knowledge about winds and currents helped them to cross short distances, like e.g.~in the Mediterranean.

Soon, the navigators realised that celestial objects, especially stars, can be used to keep the course of a ship. Such skills have been mentioned in early literature like Homer’s Odyssey which is believed to date back to the 8th century BCE. There are accounts of the ancient people of the Phoenicians who were able to even leave the Mediterranean and ventured on voyages to the British coast and even several hundred miles south along the African coast \citep{johnson_history_2009}. A very notable and well documented long distance voyage has been passed on by ancient authors and scholars like Strabo, Pliny and Diodorus of Sicily. It is the voyage of Pytheas, a Greek astronomer, geographer and explorer from Marseille who around 300 BCE apparently left the Mediterranean by passing Gibraltar and made it up north until the British Isles and beyond the Arctic Circle, where he possibly reached Iceland or the Faroe Islands that he called Thule \citep{baker_ancient_1997,cunliffe_extraordinary_2003}. Pytheas already used a gnomon or a sundial, which allowed him to determine his latitude and measure the time during his voyage \citep{nansen_northern_1911}.

\subsection{Sailing along latitude}
At these times, the technique of sailing along a parallel (of the equator) or latitude was used by observing
\newglossaryentry{circumpolar}
{
         name = {Circumpolar},
  description = {Property of celestial objects that never set below the horizon.}
}
circumpolar stars. The concept of latitudes in the sense of angular distances from the equator was probably not known. However, it was soon realised that when looking at the night sky, some stars within a certain radius around the celestial poles never set; they are circumpolar. When sailing north or south, sailors observe that the position of the celestial pole change as well, and with it the circumpolar radius. Therefore, whenever navigators see the same star
\newglossaryentry{culminate}
{
         name = {Culmination},
  description = {Passing the meridian of celestial objects. These objects attain their highest or lowest elevation there.}
}
culminating -- transiting the meridian -- at the same elevation, they stay on the ``latitude''.  For them, it was sufficient to realise the connection between the elevation of stars and their course. Navigators had navigational documents that listed seaports together with the elevation of known stars. To reach the port, they simply sailed north or south until they reached the corresponding latitude and then continued west or east.

Nowadays, the easiest way to determine one's own latitude on Earth is to measure the elevation of the North Star, Polaris, as a proxy for the true celestial North Pole. In our era, Polaris is less than a degree off. Due to the
\newglossaryentry{precess}
{
         name = {Precession},
  description = {Besides the rotation of a spinning body, the rotation axis often also moves in space. This is called precession. As a result, the rotation axis constantly changes its orientation and points to different points in space. The full cycle of the precession of the Earth's axis takes roughly 26,000 years.}
}
precession of the Earth's axis, 1000 years ago, it was $8\degr$ away from the celestial pole.

\begin{figure}[!ht]
 \resizebox{\hsize}{!}{\includegraphics{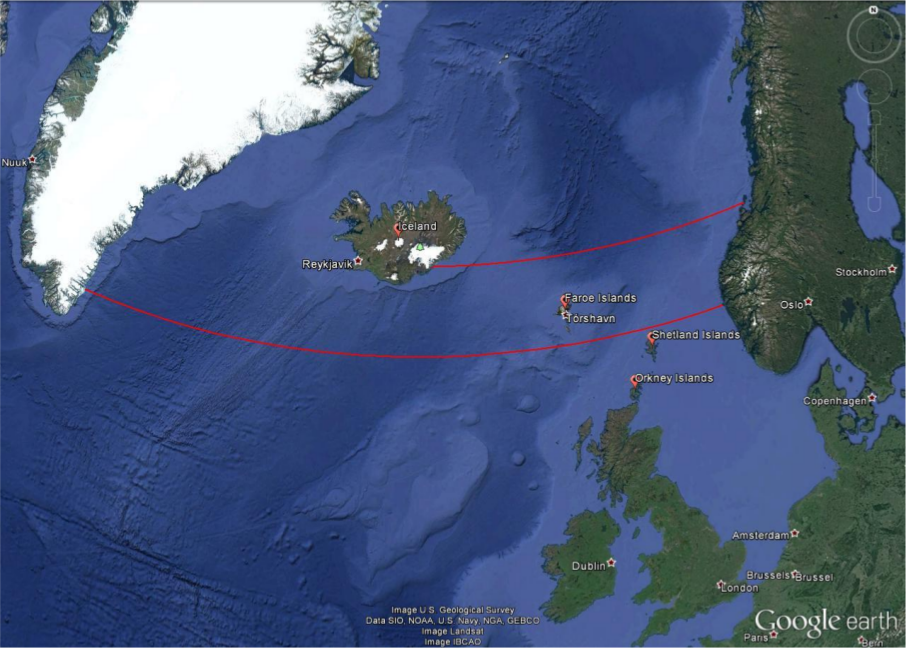}}
 \caption{Vikings probably used the technique of sailing along latitude to reach destinations west of Scandinavia (red lines). Iceland is on the 64th northern latitude and 680 nautical miles away from Norway's coast. The voyage to Greenland along the 61st northern latitude passes the Shetland and Faroe Islands. A stopover in Iceland is a viable alternative.}
   \label{f:sailing}
\end{figure}

However, using Polaris to determine the north bearing and one's own latitude of course only works when it is dark enough to see the 2~mag bright star. At a clear day, this is only possible during Nautical or Astronomical Twilight, i.e. when the Sun has set and its centre is more than $6\degr$ below the horizon. However, at latitudes higher than $61\degr$ north, the Sun can stay above such low (negative) elevations, especially around summer solstice. This is the realm north of the Shetland Islands, i.e. certainly the Faroe Islands and Iceland. Hence, observing Polaris becomes rather difficult during summer, which is the preferred season for sailing. For latitudes north of the Arctic Circle, where sea ice can block passages during winter, the sun never sets for a certain period during summer. Therefore, other techniques were needed for navigation.

\subsection{The Vikings}
\begin{figure}[!ht]
 \resizebox{\hsize}{!}{\includegraphics{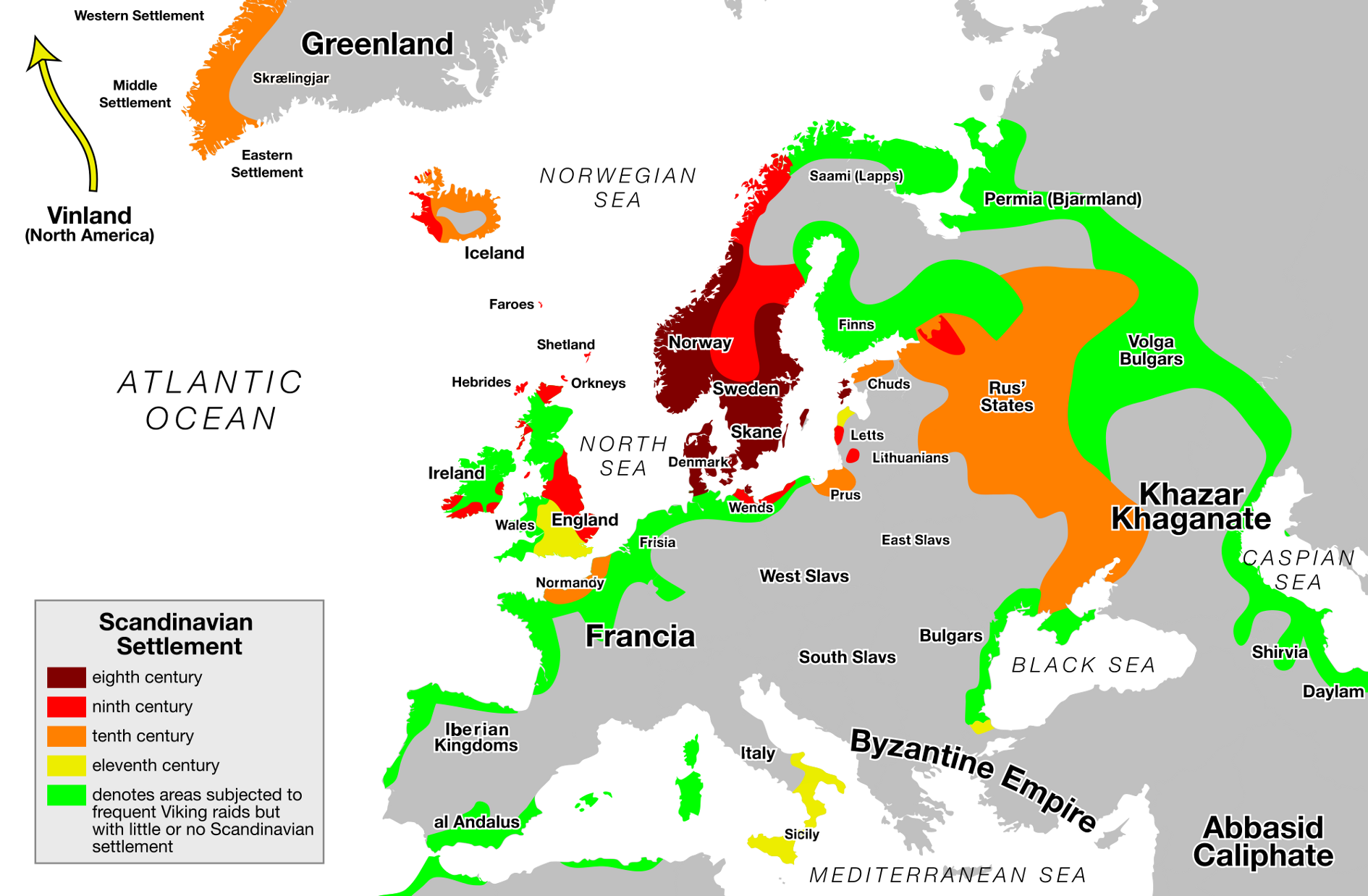}}
 \caption{Map of Viking expansion between the 8th and 11th century. Their origins are the Norwegian coast, Southern Sweden as well as Denmark (Max Naylor, \url{https://commons.wikimedia.org/wiki/File:Viking_Expansion.svg}, public domain).}
   \label{f:vikingrealm}
\end{figure}

he Vikings were Northern Germanic tribes who were known for their seamanship, their influential culture and a wide trade network. And they were feared for their raids and pillages that were executed with roaring brutality. However, contrary to common urban legends, the Vikings were not the filthy, savage barbarians that wore horned helmets when going into battle. Instead, they seemed to be well groomed, and bathed at least once per week \citep{berg_petersen_what_2012}. Their origins are the coastal regions around western and southern Scandinavia as well as Denmark. During their explorations, they settled in Iceland, Greenland, Normandy and the British Isles. However, they ventured as far as Northern America, all around Europe, the Black Sea and the Caspian Sea (Fig.~\ref{f:vikingrealm}).

\begin{figure}[!ht]
 \resizebox{\hsize}{!}{\includegraphics{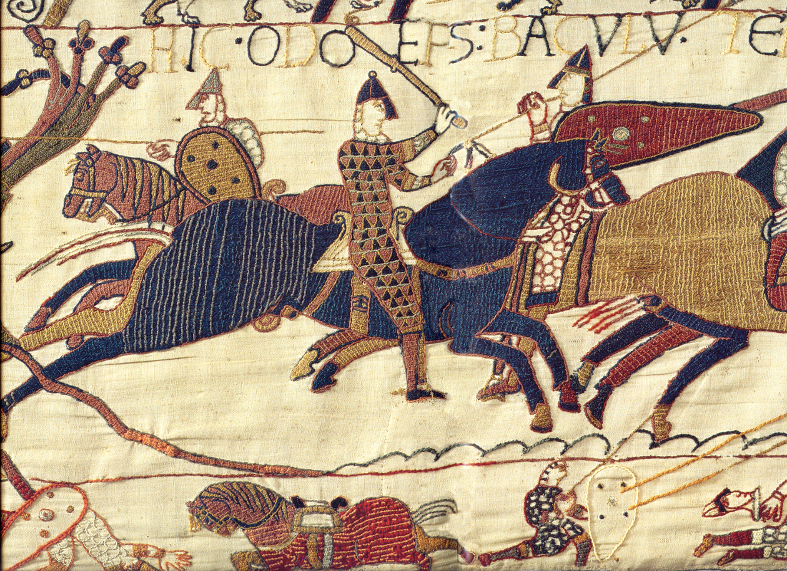}}
 \caption{A segment of the Bayeux Tapestry depicting Odo, Bishop of Bayeux and half-brother to William the Conqueror, rallying the Norman troops during the Battle of Hastings in 1066. The Bayeux Tapestry is a 70~metre long embroidered cloth depicting the Battle of Hastings and the events leading up to the Norman Conquest of England. It was probably commissioned by Odo himself \citep{hicks_bayeux_2007} (\url{https://commons.wikimedia.org/wiki/File:Odo_bayeux_tapestry.png}, public domain).}
   \label{f:bayeux}
\end{figure}

The beginning of the Viking Era is commonly dated to 793~CE with the raid of Christian monastery of Lindisfarne \citep{graham-campbell_viking_2001} in Northumbria, England. However, the Gallo-Roman historian St.~Gregory of Tours reports on an earlier attack by a Danish king named Chlochilaicus on Austrasia, the homeland of the Merovingian Franks, around 520~CE. It is believed that this Danish king may be identical to the mythical character of Hygelac in the Beowulf poem \citep{susanek_hygelac_2000}. The Viking Era ended with the Battle of Hastings between the English and the Norman-French, who were descendants of the Vikings, and the destruction and abandonment of Hedeby, an important Viking settlement and trading post, both in 1066. As it marked the Norman conquest of Britain, the Battle of Hastings was such an important turning point in British history that it was documented with colourful and vivid pictures on the Bayeux Tapestry (Fig.~\ref{f:bayeux}), made in the 1070s which is still in brilliant shape \citep{hicks_bayeux_2007}.

\subsection{Viking navigation}
The Vikings were famous for their longships, multi-purpose ships that could be used on rivers, shallow coastal waters and oceans. They were used for trade, exploration and warfare. Depending on their size, they could carry from a dozen up to 80 sailors. Because of their shallow draught, many of them did not need a harbour to make landfall but could simply be beached. Those ships were usually decorated with carved ornaments. Propulsion was provided by sail or oars which could lead to speeds of 15 to 20 knots.

\begin{figure}[!ht]
 \resizebox{\hsize}{!}{\includegraphics{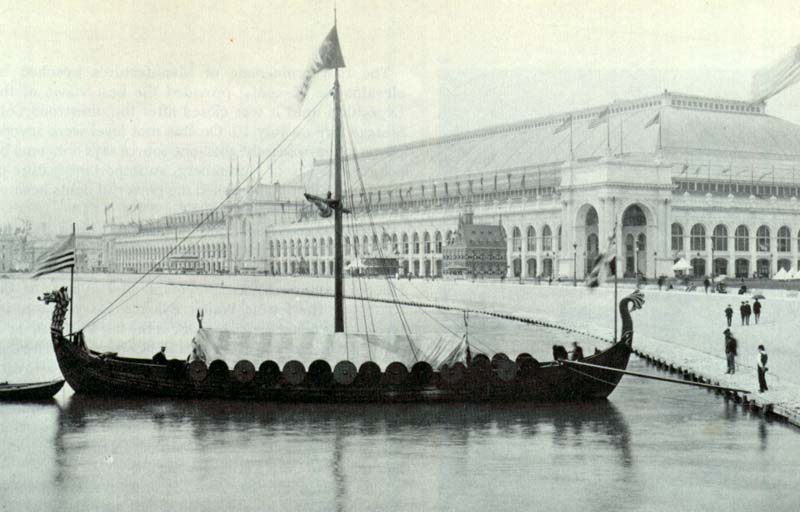}}
 \caption{The ``Viking'', a replica of the Gokstad Viking ship, at the Chicago World Fair 1893 \citep{di_cola_images_2012}. With a crew of 11, it crossed the Atlantic and reached Chicago within 2 months (public domain).}
   \label{f:gokstad}
\end{figure}

Assuming an average speed of 5 knots, crossing the Northern Sea would have been possible within one or two days. Longer trips, e.g.~from Norway to Iceland would have been achieved within five to seven days.

The Viking sailors were very experienced in interpreting the signs nature provides. They were able to read the migratory routes of birds \citep{forte_viking_2005} and whales as well as the smell and sound that the wind carries from distant shores. The Vikings probably did not have any sea charts, but they used chants and rhymes that contained sailing information as mentioned in the medieval Hauksb\'{o}k chronicle \citep{sawyer_oxford_1997} and were passed on from generation to generation. For instance, the route from southern Norway to Greenland passes the Shetland Islands and Iceland. Their sightings could be used to correct the course, which perfectly coincides with staying on latitude $61\degr$ north. Therefore, the Vikings must have had skills to follow it.

\begin{figure}[!ht]
 \resizebox{\hsize}{!}{\includegraphics{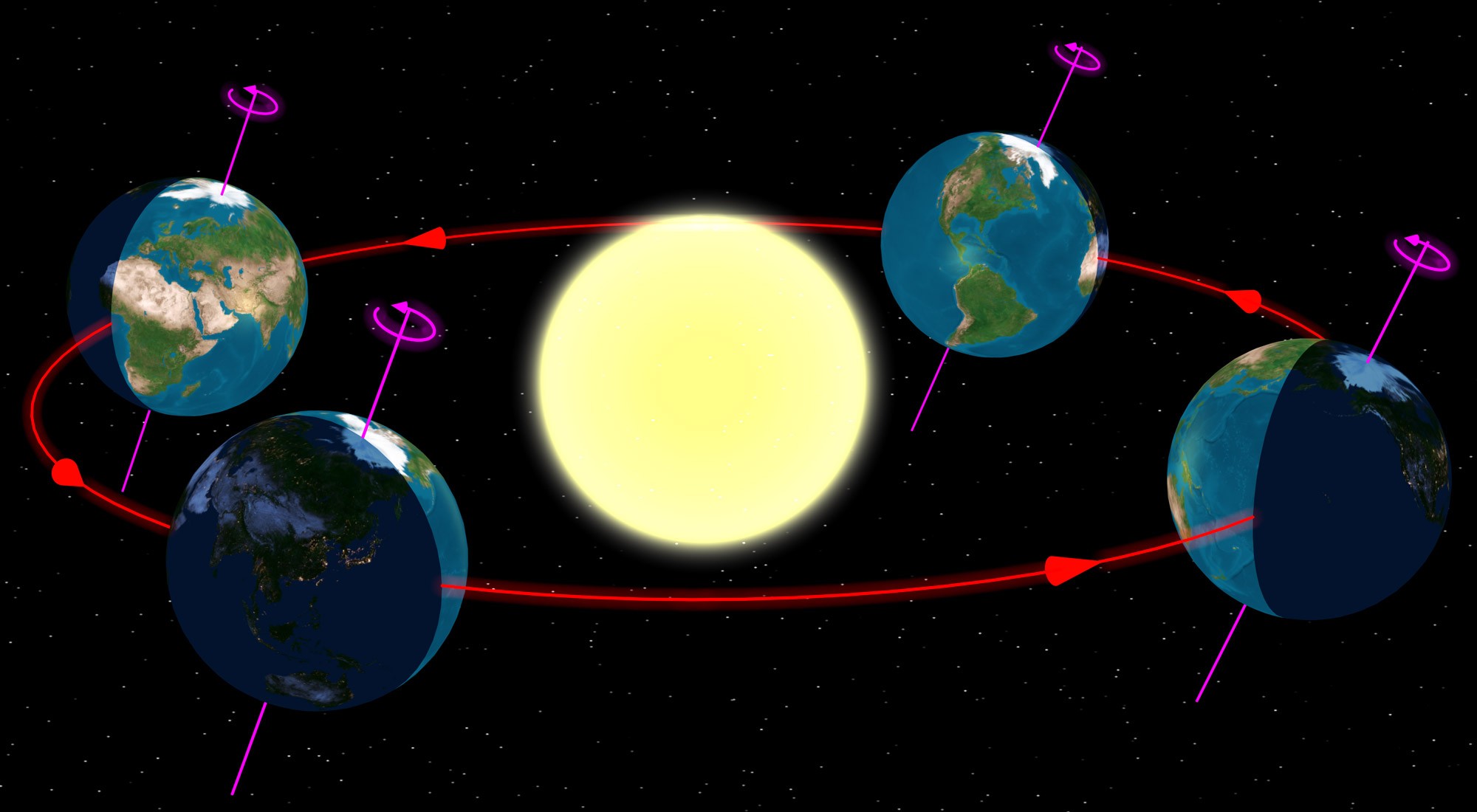}}
 \caption{Illumination of the northern and southern hemisphere of the Earth during its orbit around the Sun (Tau'olunga, \url{https://en.wikipedia.org/wiki/File:North_season.jpg}, CC0).}
   \label{f:season}
\end{figure}

As mentioned before, the Sun played an important role for finding a ship's course. The difficulty with the Sun compared to the stars is that the Sun changes its declination, i.e. the elevation above the equator. The reason is that the Earth with its tilted axis revolves around the Sun. In the northern summer, the northern hemisphere faces the Sun, while during northern winter, it is the southern hemisphere. The range, under which the Sun appears in the zenith, is the latitudes between 23\fdg4 north, the Tropic of Cancer, and 23\fdg4 South, the Tropic of Capricorn. For any given location on Earth, the Sun's elevation while it transits the meridian -- the line that connects North and South at the horizon through the zenith -- changes by the same amount.

\begin{figure}[!ht]
 \resizebox{\hsize}{!}{\includegraphics{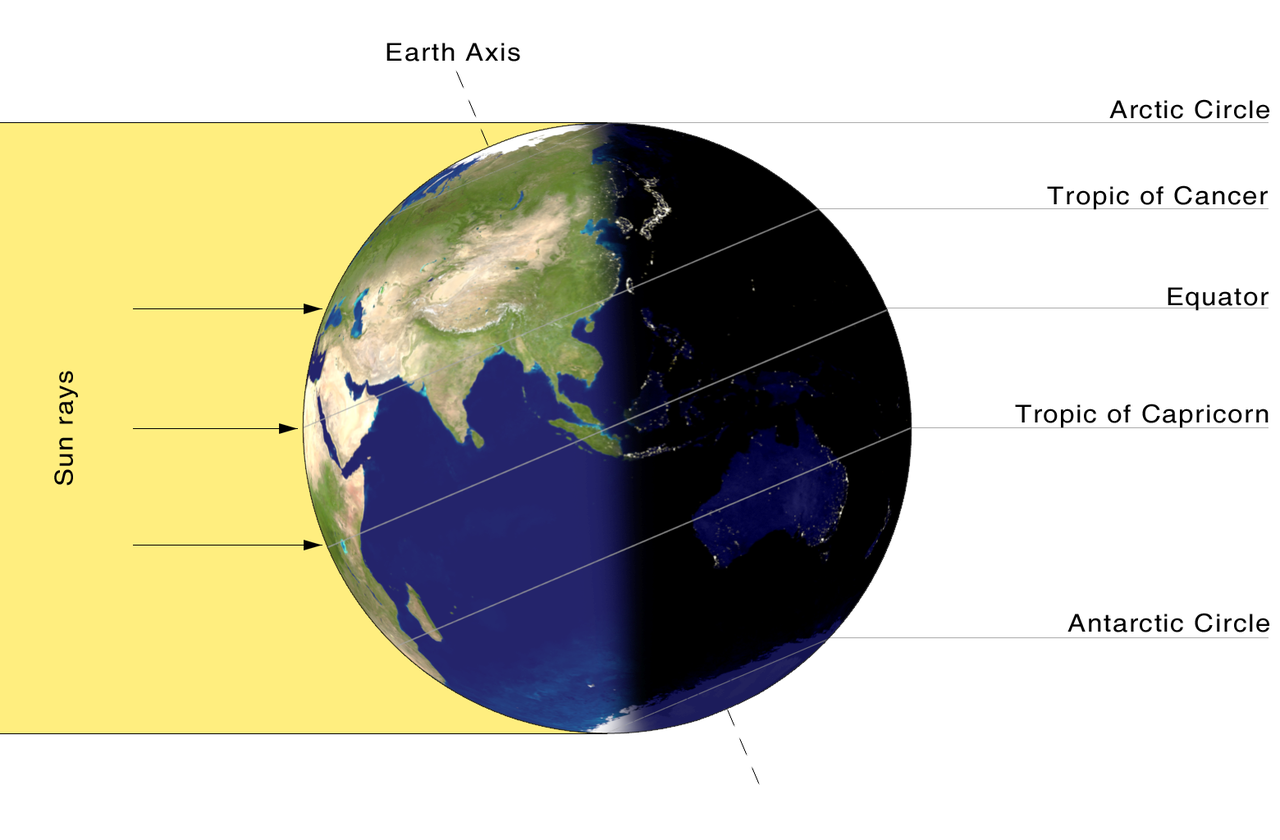}}
 \caption{At summer solstice, the Sun is directly above the Tropic of Cancer. Its apparent position changes during the year (Przemyslaw "Blueshade" Idzkiewicz, \url{https://commons.wikimedia.org/wiki/File:Earth-lighting-summer-solstice_EN.png}, ``Earth-lighting-summer-solstice EN'', \url{https://creativecommons.org/licenses/by-sa/2.0/legalcode}).}
 \label{f:solstice}
\end{figure}

\begin{figure}[!ht]
 \resizebox{\hsize}{!}{\includegraphics{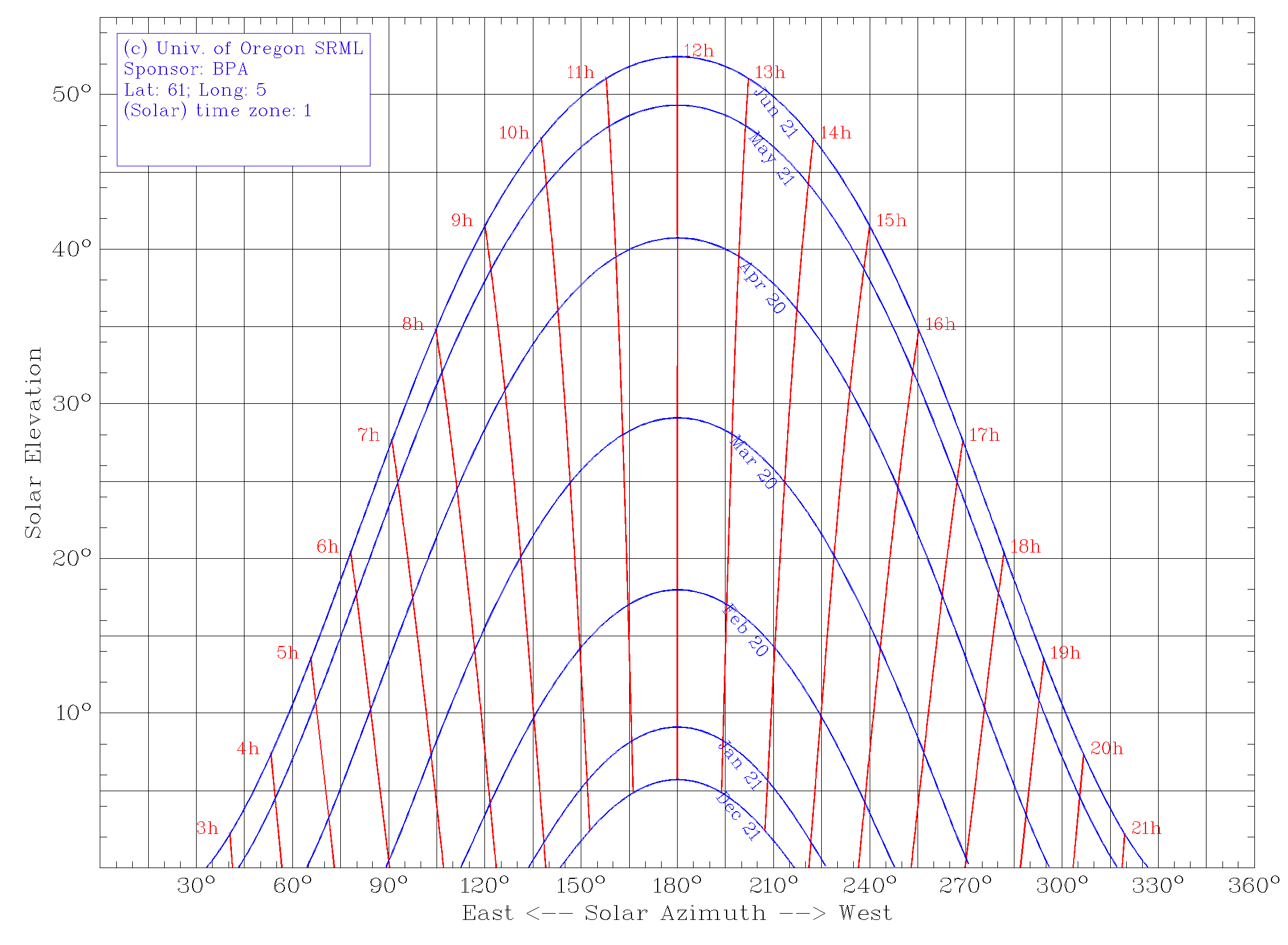}}
 \caption{The diurnal and annual elevation of the Sun above the horizon for a latitude of $61\degr$ north (Created with the sun chart path program of the University of Oregon, USA, \url{http://solardat.uoregon.edu/SunChartProgram.html}).}
 \label{f:solarpath}
\end{figure}

For latitude of $61\degr$ north, the elevation of the Sun above the horizon is shown in Fig.~\ref{f:solarpath}. South is at the centre at an azimuth of $180\degr$. Throughout the year, the elevation of the Sun at local noon changes by almost $47\degr$. However, the rate of change is not constant. We can assume a variation of the solar declination of up to $1\degr$ per voyage to be acceptable for navigational purposes.

If we allow such a variation during two consecutive days, the Sun could be used at any time of the year. This means, within two days the declination of the Sun –- or its elevation at noon -– never changes more than $1\degr$. This corresponds to a deviation of 8~km after travelling for 240 nautical miles. As already pointed out, two days are sufficient for voyages through the Northern Sea. For five to seven day journeys, the allowed period is between end of May and beginning of July. This is enough to travel from southern Norway to Iceland. The corresponding drift due to the changing solar elevation amounts to 25~km or less. For longer travels, e.g.~to Greenland, the course can be adjusted by landmarks on the way when for instance passing the Shetland Islands, the Faroe Islands and Iceland. Is there evidence for navigational tools the Vikings used with the Sun?

\newglossaryentry{diurnal}
{
         name = {Diurnal},
  description = {Concerning a period that is caused by the daily rotation of the earth around its axis.}
}

\subsection{The Sun Shadow Board}
\begin{figure}[!ht]
\centering
 \resizebox{0.5\hsize}{!}{\includegraphics{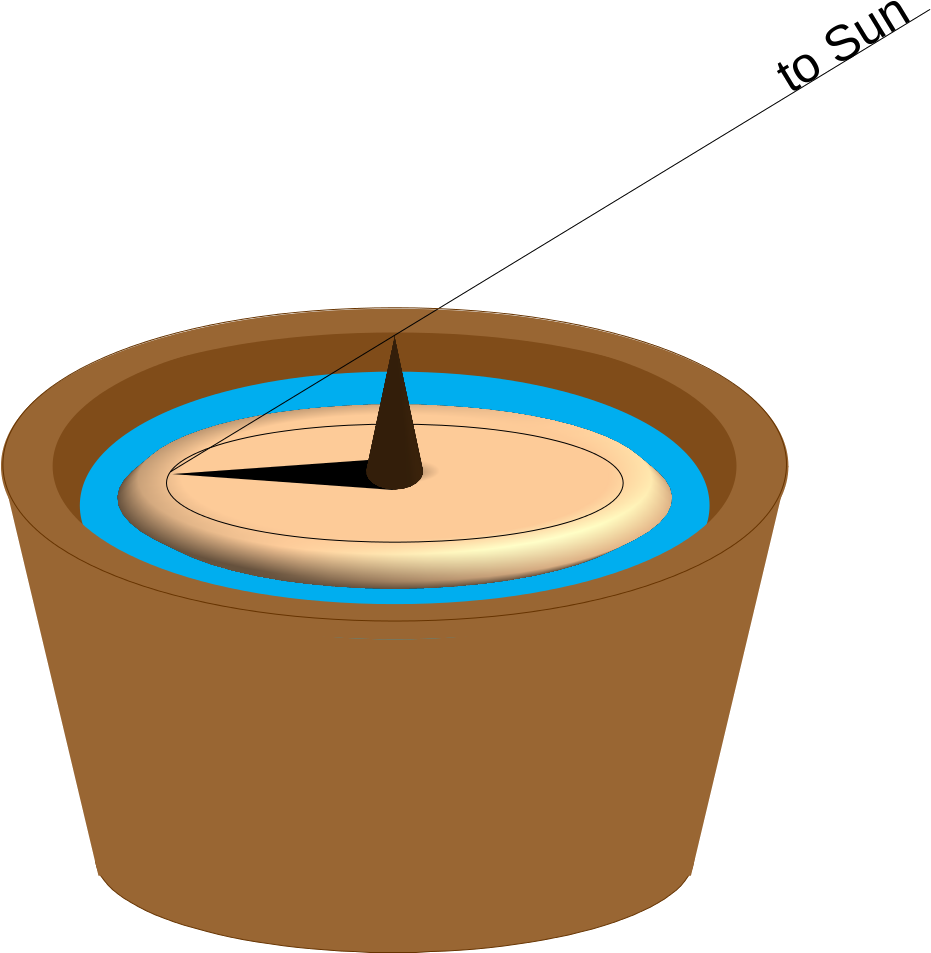}}
 \caption{Sketch of the Viking sun shadow board (M. Nielbock, own work).}
 \label{f:shadow}
\end{figure}

Sailing along latitude was probably facilitated by a device that was called {\em solskuggerfj{\o}l} (sun shadow board, Fig.~\ref{f:shadow}). 18th century sailors of the Faroe Islands have been seen using a wooden disk of up to 30~cm in diameter with engraved concentric rings and a central
\newglossaryentry{gnomon}
{
         name = {Gnomon},
  description = {Any object that casts a shadow.}
}
gnomon whose height could be adjusted \citep{tjgaard_windjamming_2011}. It was put inside a bucket with water to cancel out ship movements. It is quite likely that it was already used during the Viking age.

If we assume a negligible change of the Sun's declination, the shadow of the gnomon at noon can be calibrated to any latitude by aligning the tip of the shadow with a circle. When read during noon of the following days, the shadow should again touch the same circle. If the shadow is shorter, the position is too far South; if it is longer, the location is too far north.

\subsection{The Sun Compass}
\begin{figure}[!ht]
\centering
 \resizebox{\hsize}{!}{\includegraphics{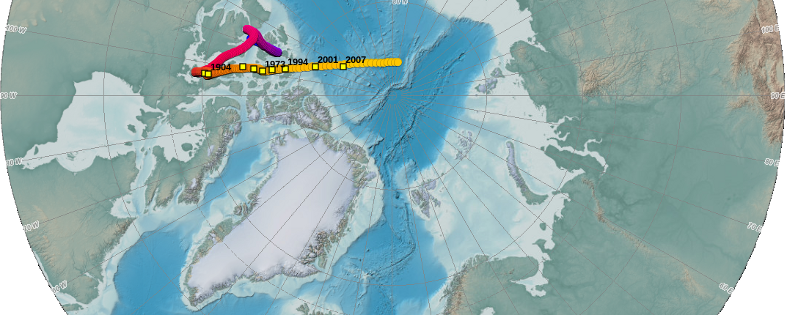}}
 \caption{Change of the magnetic North Pole and its position relative to Scandinavia (Created with the NOAA Historical Magnetic Declination Viewer, \url{http://maps.ngdc.noaa.gov/viewers/historical_declination/}).}
 \label{f:magnetic}
\end{figure}

The magnetic compass was unknown in Europe during the Viking age. And it would have been quite useless for them anyway, because the magnetic field of the Earth is far from homogeneous. The phenomenon that the magnetic poles do not align well with the geographical ones is called magnetic declination. In addition, the field lines are strongly curved. And both processes change in time (Fig.~\ref{f:magnetic}). Measuring campaigns like ESA’s SWARM satellites constantly monitor the magnetic field \citep{esa_esas_2013}.

Thus, especially at high latitudes, a magnetic compass would have let the Vikings lose their way more often than it would have aided them in finding the correct course. But it seems they were able to find the cardinal directions using the Sun.

\begin{figure}[!ht]
\centering
 \resizebox{\hsize}{!}{\includegraphics{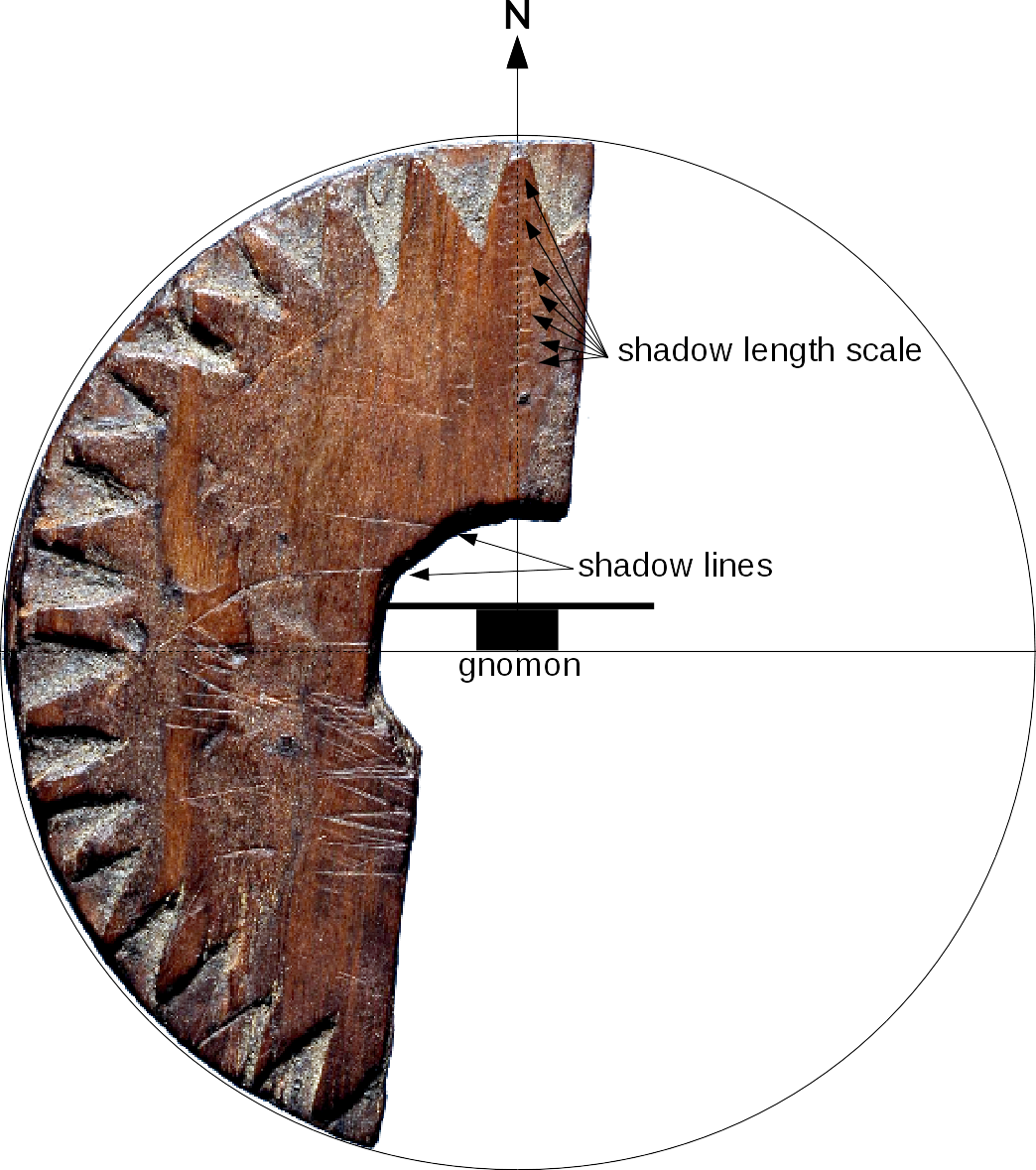}}
 \caption{Image of the original wooden disk fragment found in Uunartoq, Greenland. Annotations denote elements for its possible usage as a sun compass \citep{bernath_alternative_2013}. When the shadow is aligned with the shadow lines, North is up. Incisions in that direction permit measuring the shadow length (Lennart Larsen, Danish National Museum, \url{http://samlinger.natmus.dk/DO/10775}, ``Tr{\ae}disk\_Gr{\o}nland'', background of photograph removed and annotations added by Markus Nielbock, \url{https://creativecommons.org/licenses/by-sa/2.0/legalcode}).}
 \label{f:suncompass}
\end{figure}

In 1948, fragments of a small wooden disk were found during historic excavations of an abandoned monastery of Uunartoq in southwest Greenland (Fig.~\ref{f:suncompass}). In the following decades, people started to interpret it as a navigational tool to determine the cardinal directions using the Sun \citep{thirslund_viking_2007}. However, even to date, there are doubts that it truly served that purpose. Nonetheless, there are quite remarkable scientific analyses that demonstrate that in fact this disk could have been a combination of a
\newglossaryentry{sundial}
{
         name = {Sundial},
  description = {A stick that projects a shadow cast by the sun, i.e.~ a gnomon. The orientation and length of the shadow permits determining time and latitude.}
}
sundial, a compass, and a sun shadow board \citep{bernath_alternative_2013}.

Incised lines have been identified as paths of a shadow cast by a central gnomon during the days of
\newglossaryentry{equinox}
{
         name = {Equinox},
  description = {This is the configuration, when the Sun apparently crosses the equator. This happens twice a year. At these dates, the Sun is exactly at zenith at the Earth equator. These two dates define the beginning of spring and autumn.}
}
equinox and summer
\newglossaryentry{solstice}
{
         name = {Solstice},
  description = {This is the configuration, when the Sun apparently touches the tropics. The Sun reaches its most northern and most southern extreme. These dates define the begin of summer and winter.}
}
solstice at a latitude of $61\degr$ north. These lines hypothetically helped to determine local noon, i.e.~when the Sun attains its highest elevation when crossing the meridian. This moment indicates the time when the device can be used. At local noon, a central gnomon produces a shadow that points north. Similar to the sun shadow board mentioned above, incisions on the wooden board towards the northern direction can be used to determine possible deviations from the course along a predefined latitude.

\subsection{Local time and time zone}
The shadow of a gnomon or a sundial is shortest and points north, whenever the Sun is exactly south (northern hemisphere). This is what defines local noon. Since the Earth rotates continuously, the apparent position of the Sun changes as well. This means that at any given point in time local noon is actually defined for one longitude only. However, clocks show a different time. Among other effects, this is due to daylight saving time during summer, and the time zones (Figure 15). Here, noon happens at many longitudes simultaneously. However, it is obvious that the Sun cannot transit the meridian for all those places at the same time. Therefore, the times provided by common clocks are detached from the “natural” local time a sundial shows.

\begin{figure}[!ht]
\centering
 \resizebox{\hsize}{!}{\includegraphics{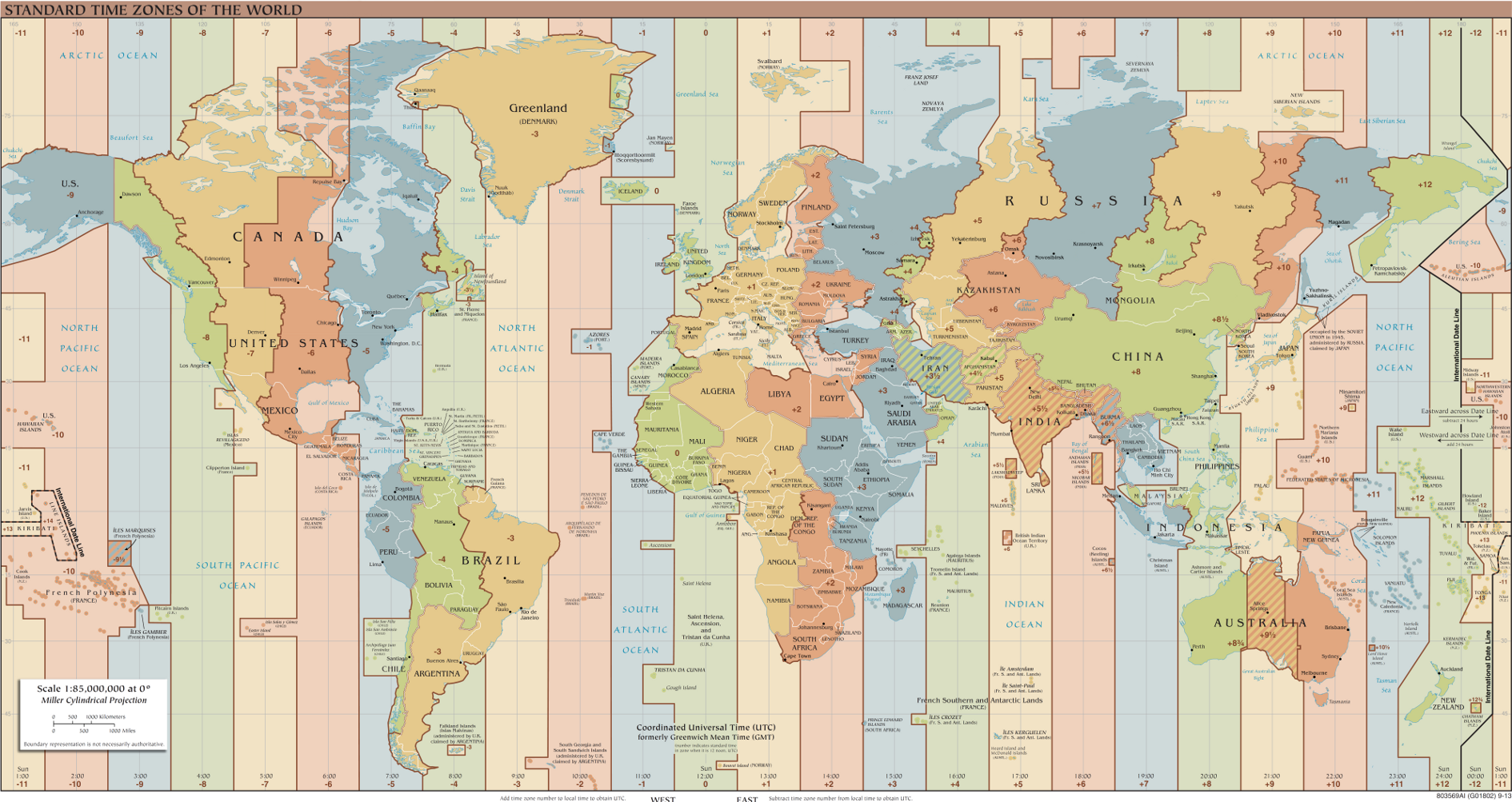}}
 \caption{World time zones. Instead of the local time that is based on the apparent course of the Sun in the sky and valid for single longitudes only, the common clocks show a time based on time zones which applies to many longitudes simultaneously (TimeZonesBoy, \url{https://commons.wikimedia.org/wiki/File:Standard_World_Time_Zones.png}, \url{https://creativecommons.org/licenses/by-sa/4.0/legalcode}).}
 \label{f:tz}
\end{figure}

\section{List of material}
The list contains items needed by one student. Some of them can be shared by two to four individuals.

\begin{itemize}
\item Worksheet
\item Cardboard (diameter at least 4 cm)
\item Toothpick
\item Earth globe (+ stable mounting, e.g. inflatable: \url{https://goo.gl/gq1d5N}, \url{http://www.unawe.org/earthball}) 
\item Compass (drawing tool)
\item Lamp or spot light
\item Scissors
\item Cutter or sharp knife
\item Glue
\item Blu Tack (or similar)
\end{itemize}

\section{Goals}
With this activity, the students will learn that
\begin{itemize}
\item the Vikings were much more than warriors.
\item the Vikings were a people that have been very influential for European history.
\item the Vikings were skilled sailors and used to navigate by the Sun.
\item the cardinal directions can be determined by the position of the Sun.
\item the astronomical (local) noon does not coincide with noon on the clock.
\end{itemize}

\newglossaryentry{cardinal}
{
         name = {Cardinal directions},
  description = {Main directions, i.e. North, South, West, East}
}

\section{Learning objectives}
At age 14, students should already understand the concepts of latitude and longitude for defining a position on Earth.

\begin{itemize}
\item After reading a story about a Viking voyage, the students will be able to explain how the Sun can be used to find the cardinal directions.
\item They will also be able to explain, how the Sun can be used to maintain a course along latitude.
\item With the same story, the students will be able to describe the region of origin of the Vikings and some of their typical professions.
\item With a hands-on activity, the students will be able to explain the basic functionality of a shadow board, an old Viking navigational tool.
\item More experience students will be able to explain the geometry behind the shadow board and how trigonometry of a right angle triangle helps to calculate the length of a shadow.
\end{itemize}

\section{Target group details}

\noindent
Suggested age range: 12 -- 16 years\\
Suggested school level: Middle School, Secondary School\\
Duration: 45 minutes

\section{Evaluation}
The evaluation is done twofold: by checking the results of the activities and from the answers of the students and their discussions.

\begin{itemize}
\item In particular, the teacher may ask the students, to what direction the Sun attains its highest elevation and how the cardinal directions can be derived from this.
\item The teacher may ask the students, how the elevation of the Sun at noon changes with latitude, and how this information can be used to navigate along it.
\item The teacher can ask the students why the length of the shadow cast by the Sun is indicative of the local latitude.
\item Together with the students, the teacher can develop a sketch that illustrates the geometry of a stick that produces a shadow when illuminated by the Sun. He then asks about the trigonometric functions involved in this.
\item While showing a map of Europe, the teacher may ask the students, where the Vikings originated from.
\end{itemize}

\section{Full description of the activity}
\subsection{Introduction}
It would be beneficial, if the activity be included into a larger context of seafaring, e.g. in geography, history, literature, etc.

Tip: This activity could be combined with other forms of acquiring knowledge like giving oral presentations in history, literature or geography highlighting navigation. This would prepare the field in a much more interactive way than what a teacher can achieve by summarising the facts.

Tip: There are certainly good documentaries available on Vikings and sea exploration that could be shown as an introduction. Some suggestions are:

\medskip\noindent
History Channel | The Vikings [New] HD (Duration: 1:28:22)\\\noindent
\url{https://www.youtube.com/watch?v=6mkib3lYA2s}

\medskip\noindent
Vikings, The Founders of Europe (Duration: 50:32)\\\noindent
\url{https://www.youtube.com/watch?v=jngPAIGrzJs}

\medskip\noindent
Building a Viking Ship (Duration: 1:46)\\\noindent
\url{https://www.youtube.com/watch?v=78kpzwGmBxk}

\subsection{Questions, Answers and Discussion}
Ask the students, if they had an idea for how long mankind already uses ships to cross oceans. One may point out the spread of the Homo sapiens to islands and isolated continents like Australia.

\medskip\noindent
Possible answers:\\\noindent
We know for sure that ships have been used to cross large distances already since 3,000 BCE or earlier. However, the early settlers of Australia must have found a way to cross the Oceans around 50,000 BCE

\medskip\noindent
Ask them, what could have been the benefit to try to explore the seas. Perhaps, someone knows historic cultures or peoples that were famous sailors. The teacher can support this with a few examples of ancient seafaring peoples, e.g. from the Mediterranean and the art of navigation.

\medskip\noindent
Possible answers:\\\noindent
Finding new resources and food, trade, spirit of exploration, curiosity.

\medskip\noindent
Ask the students, how they find the way to school every day. What supports their orientation to not get lost? As soon as reference points (buildings, traffic lights, bus stops, etc.) have been mentioned ask the students, how navigators were able to find their way on the seas. In early times, they used sailing directions in connection to landmarks that can be recognised. But for this, the ships would have to stay close to the coast. Lighthouses improved the situation. But what could be used as reference points at open sea? Probably the students will soon mention celestial objects like the Sun, the Moon and stars.

Let the students read the story below (separate document available).

\subsection{Story: Galmi the Viking}
``This is the day'', Galmi thinks by himself as he looks out into the bay of Vikebygd. He is a Norseman in his early thirties and has spent most of his life struggling to feed his family on his farm. His wife S\'{a}ga and his daughter Oda have endured a lot, but Galmi has made plans that could change things for the better. He came here 15 years ago with his parents who set out to find their own little piece of land and live peacefully. However, the last two winters have struck them hard leaving only little that could be spared for sowing out next spring.

Last night, Galmi told S\'{a}ga and Oda that he would go to Avaldsnes and sign on for the next raid to the land in the West. News had travelled about the riches his fellow Vikings had brought home from the realm of the Scots and the Picts. And he definitely wanted a share. ``We would be able to buy us food and seeds to survive until next summer'', he explains. Galmi packs his pouch for the hike that should take not much more than 9 hours. S\'{a}ga and Oda are sad and afraid, but they also know that this would probably be their last chance. And if everything turns out for the best, Galmi should be back within a week the latest.

He arrives in Avaldsnes just before sunset. It is a busy little town with about 15 houses, a market place, the chief's residency, and, of course, a harbour with a handful of longships anchoring. Tomorrow will be the last raid for this season, and Galmi wants to be on board. He enters a tavern and quickly finds a group of experienced sailors. They share a few drinks and Galmi befriends them. It turns out that one of them is the nephew of an old friend of his fathers, Floki. And Ragnar is the leader of this group. They are about to head for the mead hall where the chief of the island of Karm{\o}y, Augvald, assigns the raiding parties to their ships. Galmi is lucky. Together with Ragnar, Floki and 21 other Vikings, he will set sail for Inbhir Nis next morning, a rich Celtic village at the mouth of the river Ness.

Galmi does not get much sleep during the night. He is torn between the excitement of the journey and the battle ahead, and the insecure future of his family. Just as his fatigue starts to win against the racing thoughts in his head, Ragnar calls him. And Floki is already at his side. It is before sunrise and the chilling breeze from the North delivers a glimpse of the nearing autumn. As they reach the shore, three ships have already left the pier and 8 oars at each side of the boats accelerate them quickly north to the mouth of the fjord. The rest of Galmi's crew is already here and prepares the ship. It is a real beauty. ``Galmi! Here is your seat'', shouts Rollo. ``You'll be among the first to row!'' Galmi shrugs, ``Sure! Why not?'' Ragnar and Floki smile and they jump on board.

One hour later, the ship leaves the fjord and the winds turn out to be favourable. They set their sail and the ship accelerates smoothly out to the open sea. Galmi is excited and surprised how steady this ship is despite the waves and the shallow draught. It is a sunny day and the gods seem to be with them. 

The journey continues without any problem. Galmi uses this opportunity to get to know the crew. They tell him stories from their last expeditions and how proud they are being a part of this. ``Galmi, bring me a bucket with water'', Floki suddenly suggests. ``What for?'', Galmi returns puzzled. ``You’ll see.'' Galmi hands him the bucket while Floki grabs a wooden disk. ``It's noon and the Sun is shining, let's see.'' He puts it into the bucket where it begins to float. As Galmi looks closer, he discovers a ring incised about half way between the edge of the disk and the centre, where a small cone sticks out. A shadow extends to the edge of the board and slightly beyond the ring. ``Okay,'' says Floki, ``we have to continue our course to the Southwest for another five hours. Then we’ll turn west.''

Galmi is surprised. Is Floki a seer or a magician? Is he in contact with Kvasir, the god of wisdom? ``Not at all'', laughs Ragnar, ``he is just our navigator. Floki, can you explain what you do?'' “Sure. You see, Galmi. The Sun shines on this board and the shadow points north. You know that the Sun is south at noon, do you? The length of the shadow tells me, if we are on course. If the shadow is too long, we are too far north. The Sun is lower there. Is it too short, we know that we are too far south.'' ``How do you know all this?'', asks Galmi. ``My grandfather taught me. He was one of the first to scare the hell out of the Scots'', Floki replies with pride in his eyes.

Four hours after the course correction the steady breeze picks up and dark clouds appear at the horizon. ``Darn! Seems like Thor is angry again!'', Ragnar shouts. ``Get prepared for some rough weather ahead!'' The wind evolves into a storm. Floki strikes the sail. ``Okay guys! Take the oars and row!'' Ragnar has to shout in order to be heard through the howling wind. Galmi pulls as hard as he can. It starts to rain heavily and thunder and lightning seem to tear the skies apart. ``Pitch the tent!'', Ragnar shouts. The crewmen pull up a thick and long piece of linen and cover the ship almost from bow to stern. The ship rolls heavily, but the men keep on rowing. Two hours later, the crew at the oars changes. Galmi tries to get some sleep. He is so tired that neither the rocking boat nor Thor's hammer can keep him awake.

Galmi dreams of S\'{a}ga and Oda. He feels a warm sensation on his cheeks and opens his eyes. The Sun is rising and the wind has died, but through the silent hissing of the waves the snoring of the crew sounds like the howling of the Fenrir wolf. The sail is up and Floki leans at the rudder. ``How did it go, Floki?'', Galmi asks. ``Oh, it wasn’t too bad. Nothing we hadn’t managed before. It is about 12 hours until we reach the shores.'' ``Come, let me take the rudder for a while and get some rest. I think I can hold that stick for you.'' Floki smiles. ``Thank you, my friend.''

Twelve hours later Ragnar hears the distant calling of seagulls. Everyone on board knows that this means they are right on course and the shore is not very far. ``Can you smell the smoke, Galmi?'', asks Floki. ``The Scots burn peat for heating their homes.'' ``Land ahead!'' Rollo is the first to discover the coastline. But it is another four hours until Inbhir Nis. Ragnar discusses the plan with his crew. They will beach outside the town and camp until the next morning. After the tiring passage everybody needs the rest. Two days after they left Avaldsnes, Ragnar, Floki, Rollo and Galmi are prepared as is the rest of the crew. They ask for Odin's support and head for Inbhir Nis, whose clueless inhabitants will be struck by surprise. ``This is the day'', Galmi says to Floki.

\subsubsection{Questions about the story}
Q: What is Galmi’s job?\\\noindent
A: He is a farmer.

\medskip\noindent
Q: Why does he want to go on a raid?\\\noindent
A: Harvests have been poor. He needs resources to buy seeds.

\medskip\noindent
Q: Who is leading the crew of sailors that Galmi joins?\\\noindent
A: Ragnar

\medskip\noindent
Q: What is their destination? Would you know its modern name?\\\noindent
A: Inbhir Nis is the old Gaelic name of Inverness.

\medskip\noindent
Q: What propulsion do longships have?\\\noindent
A: Oars and sail

\medskip\noindent
Q: How do the Vikings protect themselves against heavy rainfall on open seas?\\\noindent
A: A linen tent

\medskip\noindent
Q: How does Floki determine the course?\\\noindent
A: He uses a shadow board to determine their latitude.

\medskip\noindent
Q: How does the crew know that the coast is near?\\\noindent
A: They observe seagulls. They smell smoke.

\medskip\noindent
Q: Do longships need anchors to dock?\\\noindent
A: No, longships are shallow enough to be beached.

\subsection{Activity: A Game of Shadows}
This activity employs the educational tool ``Parallel Earth'' (\url{http://www.unawe.org/resources/books/
Parallel_Earth/}) which demonstrates on model level the phenomena related to the Sun illuminating the Earth. The students produce a miniature version of a shadow board and experiment with it.

If there are not many globes available, the teacher may choose to introduce this activity as a demonstration with only one globe. However, there are inflatable globes available for only a few Euros that can be distributed among the students. In that case, let them work in groups of two.

\subsubsection{Building a miniature shadow board}
\begin{figure}[!ht]
\centering
\resizebox{\hsize}{!}{\includegraphics{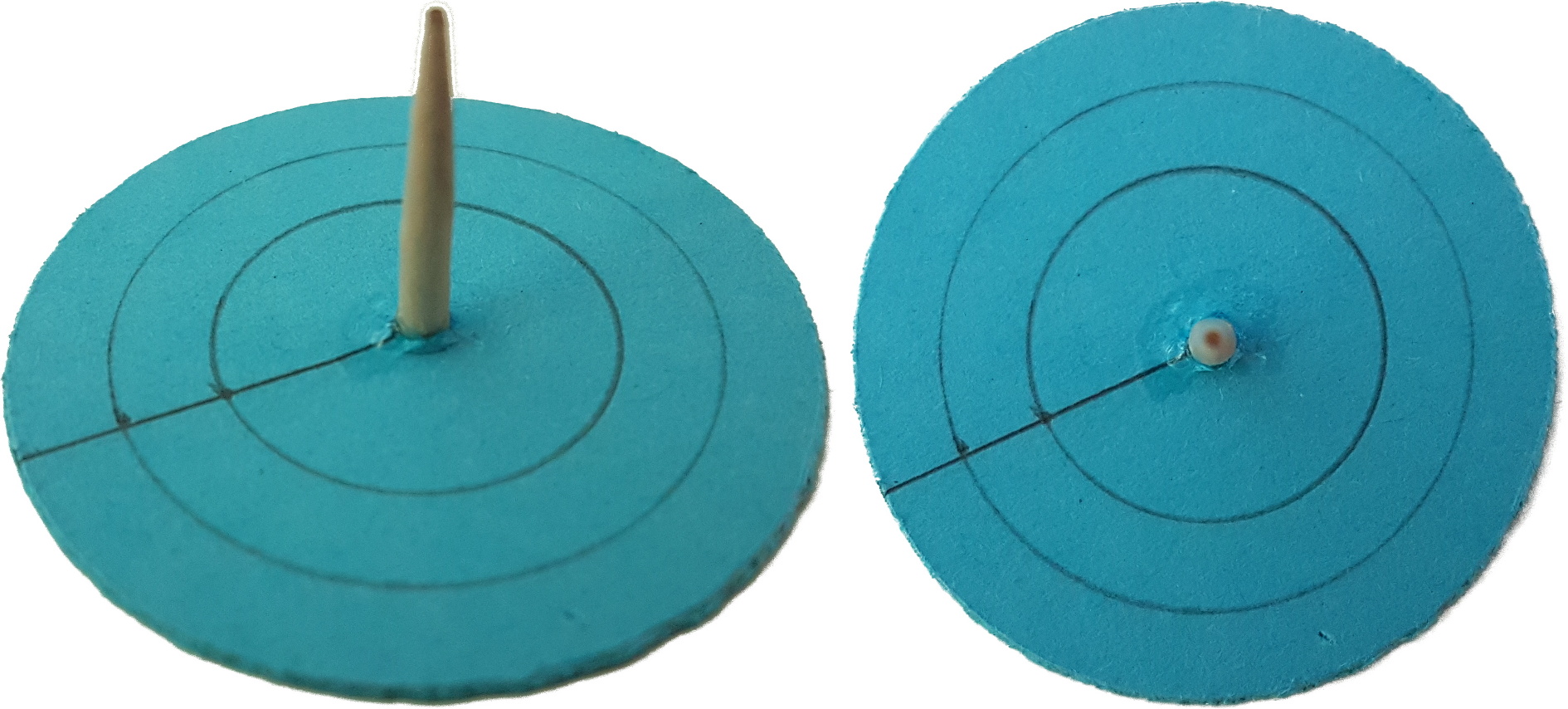}}
\caption{Miniature shadow board made of cardboard and a toothpick. Left: side view, right: top view (own work).}
\label{f:gnomon}
\end{figure}

\begin{enumerate}
\item Use the compass to draw three concentric circles with radii of 1, 1.5, and 2~cm on a piece of cardboard.
\item Cut out the disc along the circle with the radius of 2~cm.
\item Cut a 2~cm long piece from the toothpick.
\item Run it through the centre of the cardboard disc and glue it. Make sure that it remains perpendicular to the surface of the disc. The two rings on the disc must be on the same side like the stick (see Fig.~\ref{f:gnomon}).
\item Let the glue dry.
\end{enumerate}

\subsubsection{Part 1: Shadows}
Put the shadow board on the desk. Before doing some experiments, answer the following questions:

\medskip\noindent
Q: When you illuminate the shadow board, how will the shadow cast by the stick behave when changing the position of the lamp? What will be the direction of the shadow relative to the lamp?\\\noindent
A: It always points away from the lamp.

\medskip\noindent
Q: How will the shadow change when you hold the lamp high or low?\\\noindent
A: The higher the lamp, the shorter the shadow.

\medskip
Take a lamp and illuminate the shadow board. Change the position of the lamp relative to the board and observe how the shadow changes.

Compare your observations to your predictions. Did the shadow behave as you have anticipated? Discuss with your classmates:

\medskip\noindent
Q: Imagine the lamp represents the Sun. Where does the shadow point to, when the Sun is south (northern hemisphere)/north (southern hemisphere) of you?\\\noindent
A: The shadow points north (northern hemisphere)/south (southern hemisphere).

\medskip\noindent
Q: Imagine the lamp represents the Sun. Are the shadows longer in winter or summer? Explain!\\\noindent
A: As the Sun attains lower elevations above the horizon during winter, this is the season when the shadows are longer.

\subsubsection{Part 2: Navigate}
The Vikings had travelled between Scandinavia and Greenland. Fig.~\ref{f:sailing} demonstrates that when embarking from southern Norway and travelling along the 61st northern latitude, they would end up at the southern tip of Greenland. It is quite likely that at least during parts of such a voyage they used the navigational tool of the shadow board.

\begin{figure}[!ht]
\centering
\resizebox{\hsize}{!}{\includegraphics{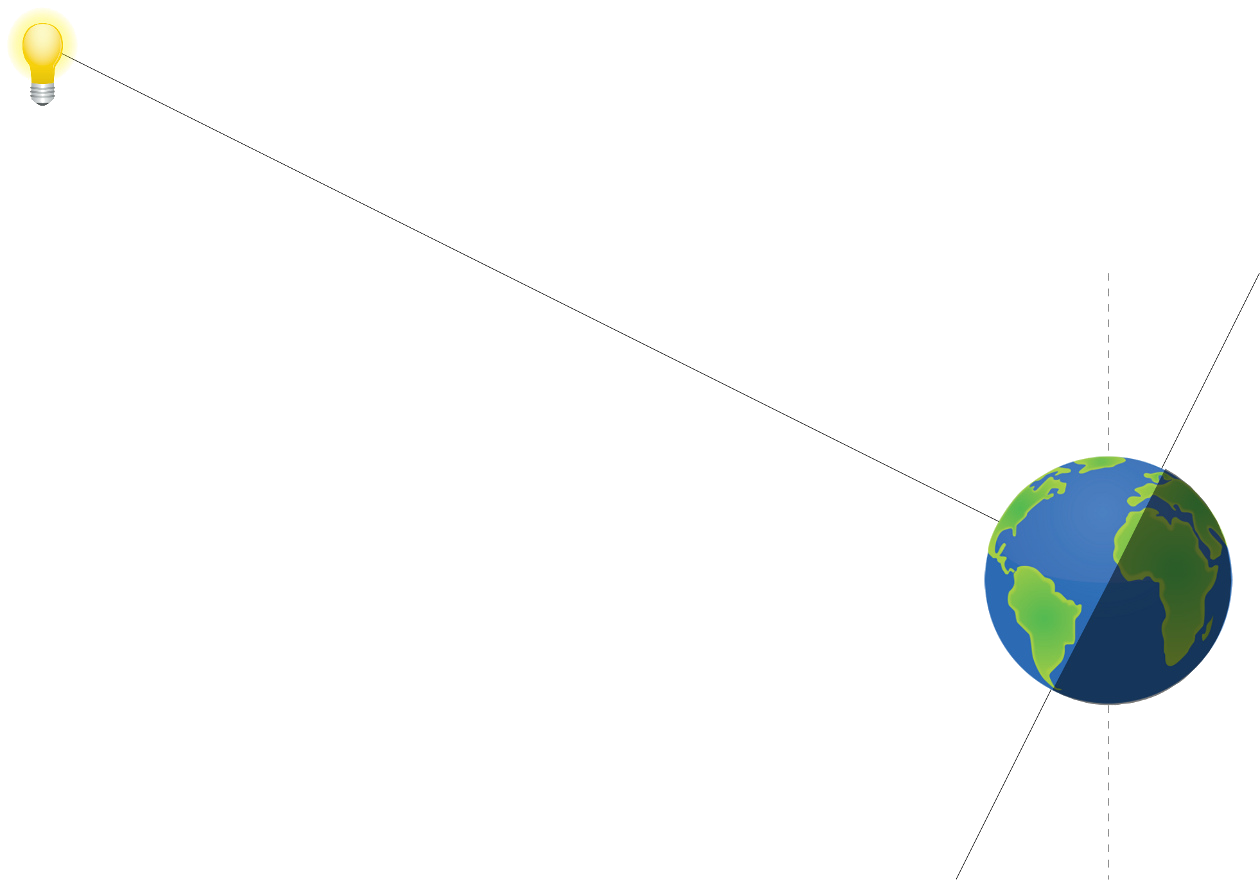}}
\caption{The lamp that represents the Sun illuminates the globe. Northern summer is achieved, when the axis of the globe and the line connecting it with the lamp subtends an angle smaller than $90\degr$ (own work).}
\label{f:summer}
\end{figure}

To simulate the point of embarkment, attach the miniature shadow board with a Blu Tack on the globe at the southern tip of Norway. Since the Vikings predominantly sailed during summer, the angle between the axis of the globe and the connecting line to the lamp should be less than a right angle ($90\degr$, see Fig.\ref{f:summer}) -- although this is not very important for the purpose of this exercise. The easiest way to do this is to keep the North Pole up.

Important: Always try to keep the lamp, the shadow board and the North Pole within the same vertical plane.

\begin{figure}[!ht]
\centering
\resizebox{\hsize}{!}{\includegraphics{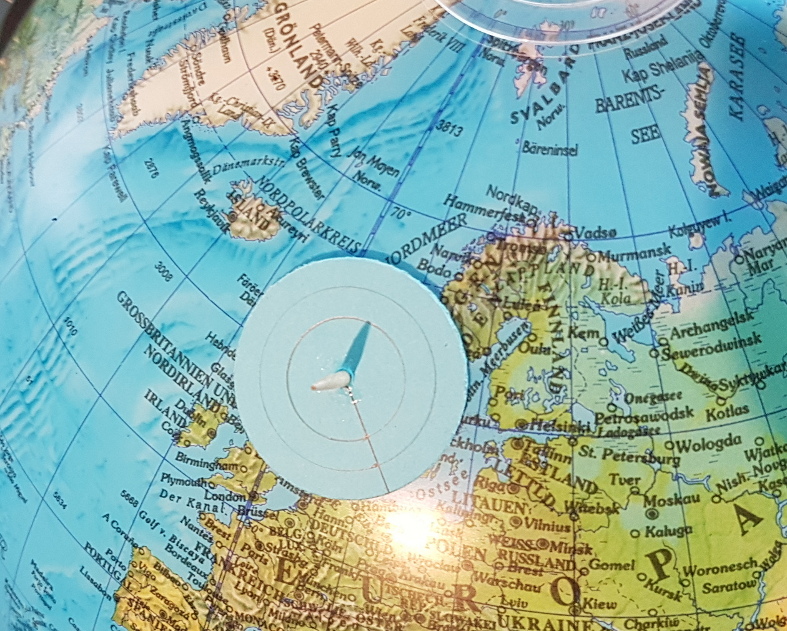}}
\caption{Starting position of the shadow board with the gnomon located at southern Norway (own work).}
\label{f:globe1}
\end{figure}

In the beginning, put the globe in a position so that the shadow on the shadow board points up, i.e.~to the North Pole. Adjust the distance between the globe and the lamp until the shadow touches one of the rings (inner ring in Fig.~\ref{f:globe1}).

\medskip\noindent
Q: What is the local time of the day, when the lamp (the Sun), the shadow board and the axis of the globe (Earth) are within the same plane? From the perspective of the shadow board on the Earth, where would be the Sun in the sky?\\\noindent
A: The Sun just passes the meridian, i.e. it is due south. Therefore, this situation is local noon.

\begin{figure}[!ht]
\centering
\resizebox{\hsize}{!}{\includegraphics{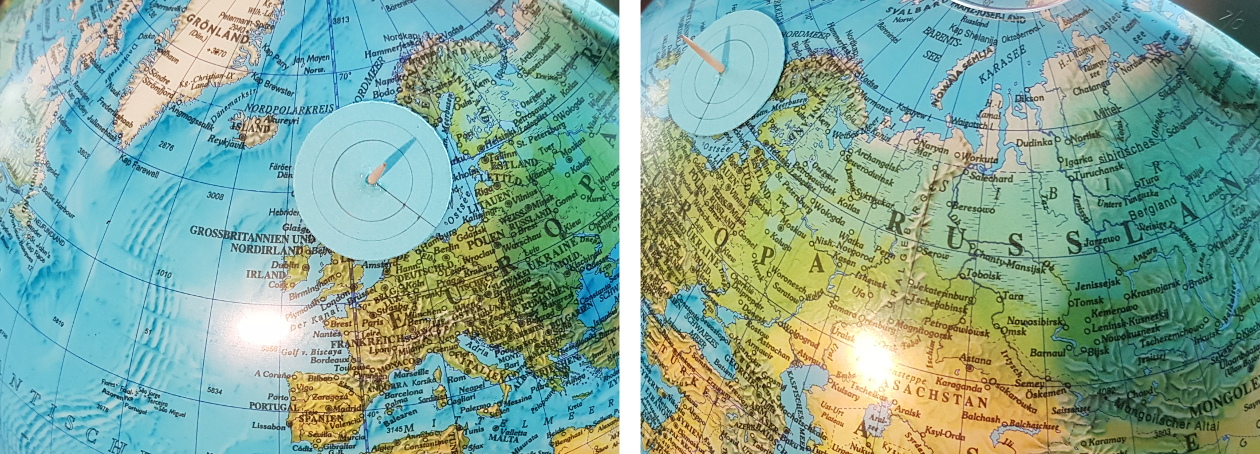}}
\caption{Change of shadow for situations before (left) and after local noon (right). The orientation of the shadow indicates time (own work, not presented to the students).}
\label{f:globe2}
\end{figure}

\medskip
While keeping the orientation of the globe, rotate it around its axis and watch how the shadow changes (Fig.~\ref{f:globe2}).

\medskip\noindent
Q: Which situation corresponds to the configuration with the shadow board east of the previous orientation of the globe?\\\noindent
A: This is the morning of the day.

\medskip\noindent
Q: In which way does the shadow change?\\\noindent
A: It turns away from the previous direction and becomes longer.

\medskip\noindent
Q: When is the shadow shortest?\\\noindent
A: At local noon.

\begin{figure}[!ht]
\centering
\resizebox{\hsize}{!}{\includegraphics{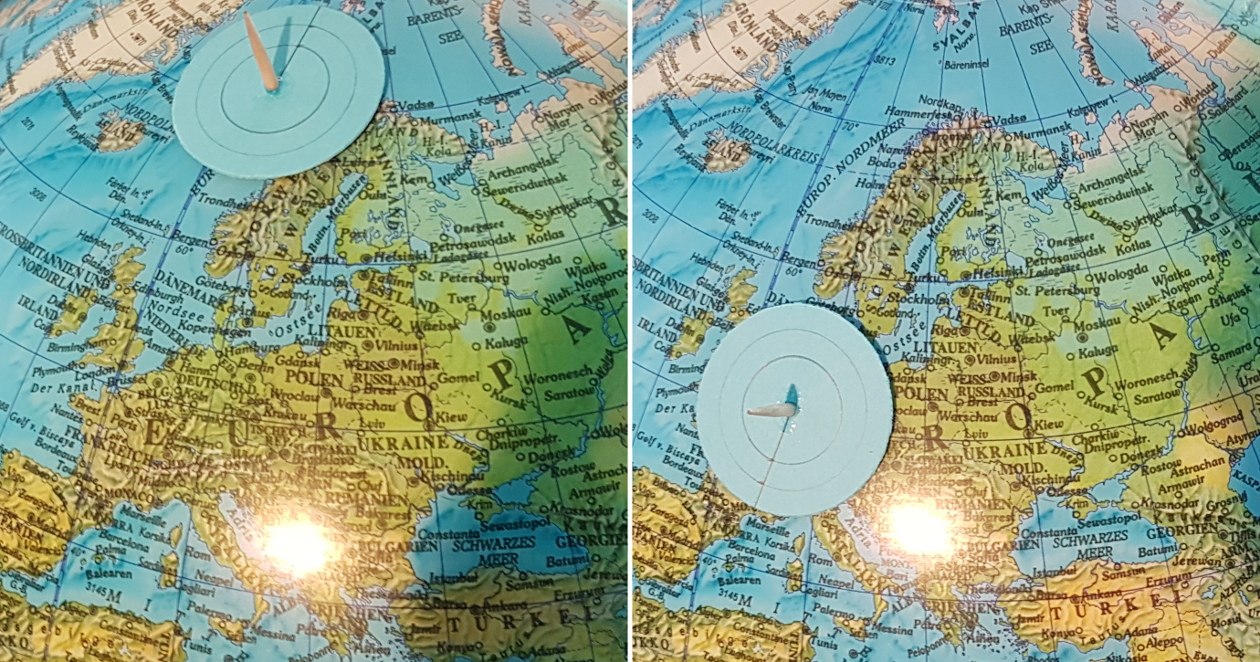}}
\caption{Change of shadow during local noon for situations more northern (left) and more southern (right) than the initial position. The length of the shadow at noon indicates latitude (own work, not presented to the students).}
\label{f:globe3}
\end{figure}

\medskip
Now, return to the initial configuration. Then change the latitude of the starting position by moving the shadow board north and south (Fig.~\ref{f:globe3}).

Tip: If an inflatable globe is used, its orientation can be fixed by mounting it on a bowl.

\medskip\noindent
Q: In which way does the shadow change?\\\noindent
A: The length changes. It is shorter for more southern positions, and longer for more northern positions.

\medskip
You will now simulate a voyage to the southern tip of Greenland. During each measurement with the shadow board adjust the globe such that it indicates local noon for the board.

Now put the shadow board on a position west along the current latitude. Rotate the globe until the shadow points north. Repeat this procedure a few times until you reach the southern tip of Greenland. Make sure, the tilt of the axis is kept constant.

\begin{figure}[!ht]
\centering
\resizebox{\hsize}{!}{\includegraphics{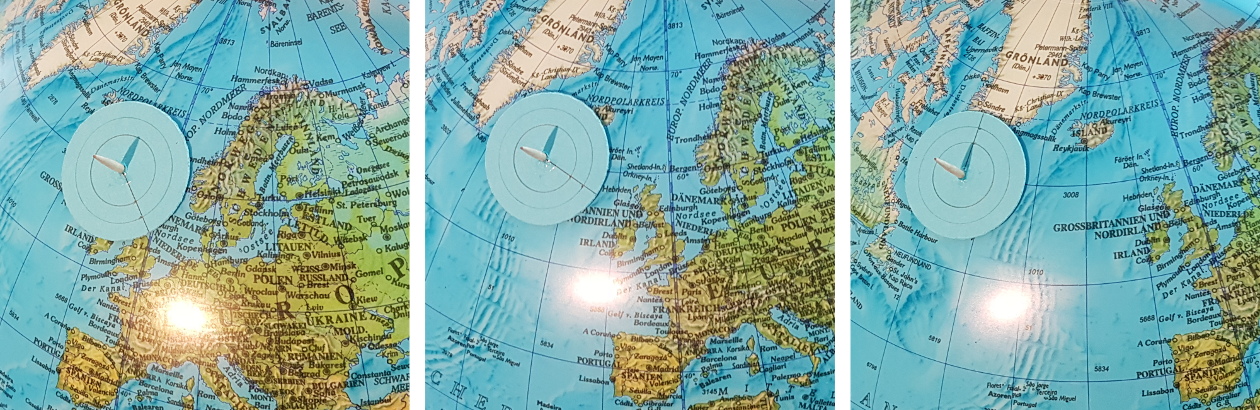}}
\caption{Functionality of the shadow board when travelling west along latitude. When measured at local noon, the orientation and the length remains the same (own work, not presented to the students).}
\label{f:globe4}
\end{figure}

\medskip\noindent
Q: Explain why the length of the shadow during local noon is always the same anywhere on the same latitude.\\\noindent
A: The configuration between the lamp (the Sun) and any position on a given latitude is always the same. Rotating the globe (Earth) does not change the angle under which the shadow board is illuminated.

\subsection{The Math of the Shadow Board (optional, for higher terms, trigonometric functions needed)}
This topic is an interesting application of simple trigonometry. Therefore, the example of the shadow board can be used to motivate reinforcing trigonometry.

For simplicity, the calculations assume:
\begin{itemize}
\item The Sun is above the equator (equinox).
\item The radius of the Earth is negligible in comparison to Sun’s distance.
\item Local noon at the location of the shadow board.
\item The local horizon at the position of the shadow board is flat.
\end{itemize}

\begin{figure}[!ht]
\centering
\resizebox{\hsize}{!}{\includegraphics{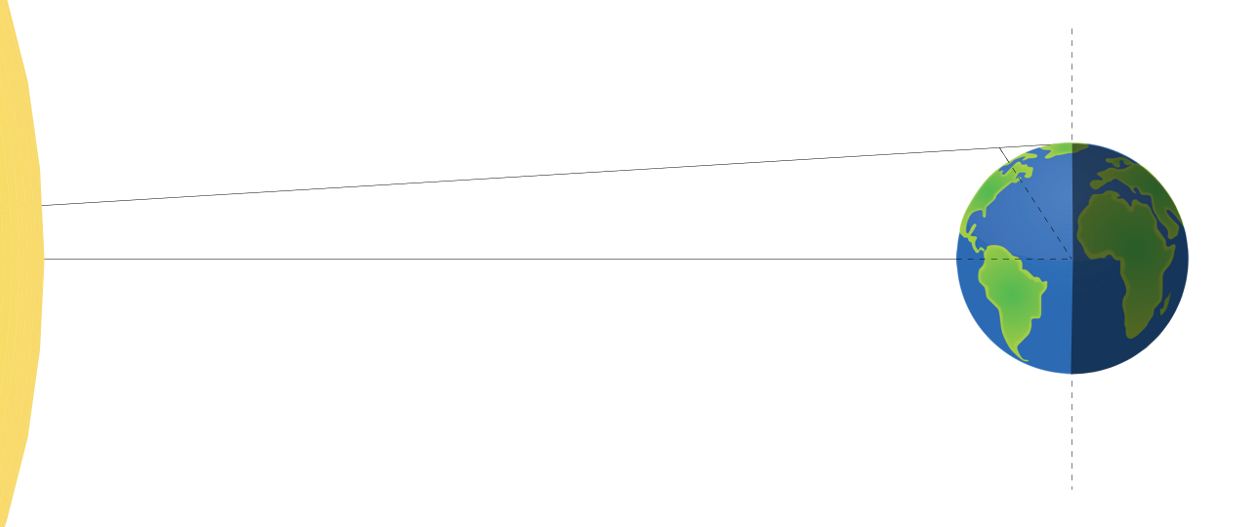}}
\caption{Sketch that represents the illumination of a gnomon at a certain latitude on Earth by the Sun. For simplicity, the Sun is located above the equator (equinox). Local noon is assumed (own work).}
\label{f:sun-earth}
\end{figure}

\begin{figure}[!ht]
\centering
\resizebox{\hsize}{!}{\includegraphics{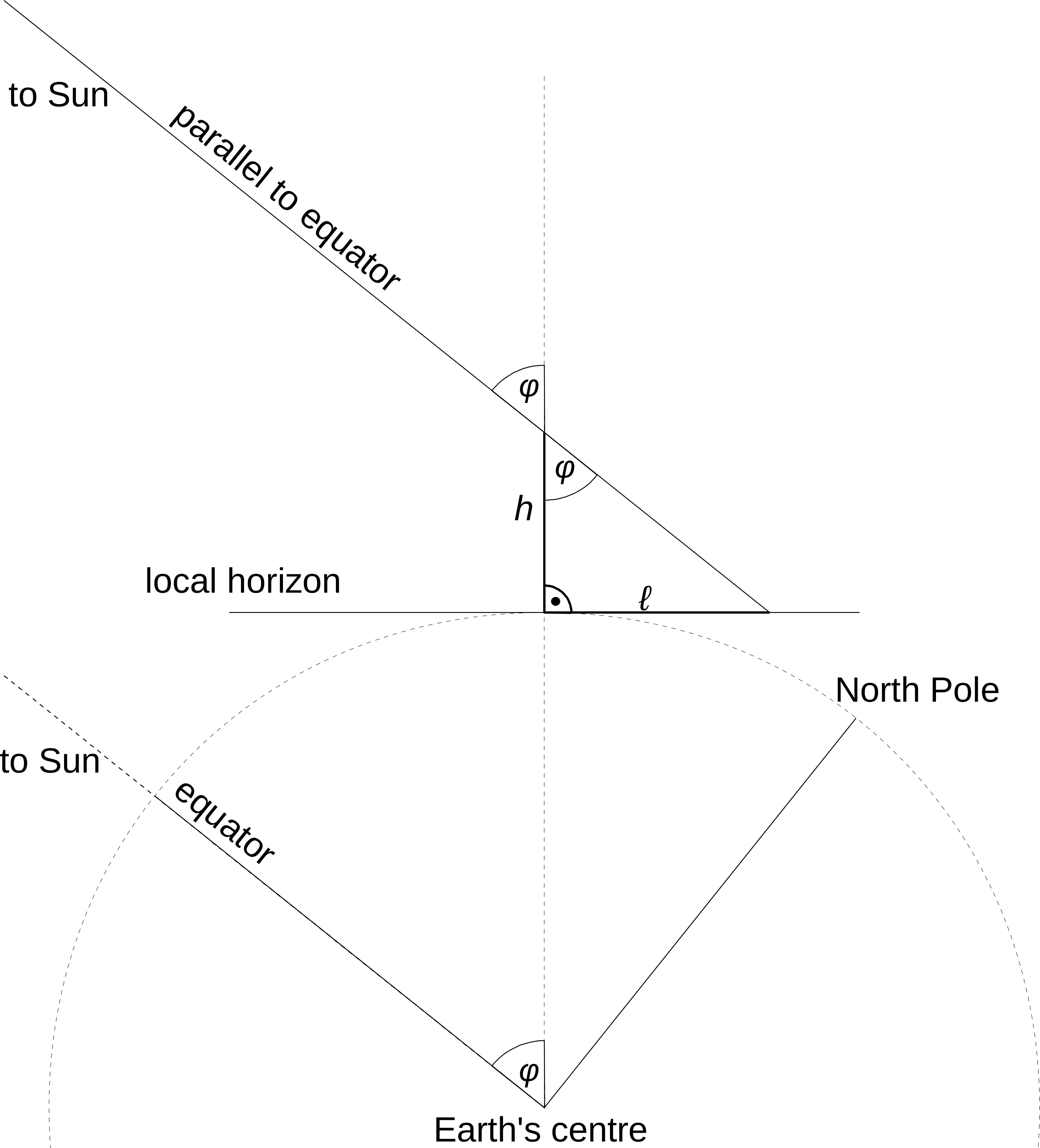}}
\caption{Sketch that demonstrates the geometric relations of a gnomon (stick) of height $h$ at latitude $\varphi$ that, illuminated by the Sun, casts a shadow of the length $\ell$. For simplicity, the Sun is located above the equator (equinox). The local horizon is assumed to be flat (own work).}
\label{f:shadowlength}
\end{figure}

The basic configuration of the Sun and Earth is shown in Fig.~\ref{f:sun-earth}. There is a gnomon (i.e.~the shadow board) extending from the surface of the Earth that, illuminated by the Sun, casts a shadow. Since the Earth is small in comparison to the Sun's distance, the rays of light hitting the equator and the gnomon are considered to be parallel to each other (Fig.~\ref{f:shadowlength}).

Since the local horizon at the position of the shadow board is a tangent at the surface of the Earth, the gnomon and the horizon subtend a right angle. The height of the gnomon is $h$ and the length of the shadow is $\ell$. The geometry of Fig.~\ref{f:shadowlength} demonstrates that the latitude $\varphi$ is also the angle opposite of $\ell$.

As a task, develop Fig.~\ref{f:shadowlength} with the entire class or group.

Our local coordinate system that extends to the horizon is a tangent area that touches the surface of the Earth. When looked at from the side, this area appears like a line that touches the circumference of the Earth -- a circle.

\medskip\noindent
Q: What is the angle between a tangent and a line connecting it to the centre of the circle?\\\noindent
A: It is a right angle.

\medskip\noindent
Let them identify the right angled triangle.

\medskip\noindent
Q: What is the trigonometric function that describes the relation between $\ell$ and $h$?\\\noindent
A: The tangent.

\medskip\noindent
Q: If $h = 10\,{\rm cm}$, what is the length of the shadow $\ell$ at latitudes of $\varphi = 40\degr$ and $60\degr$?\\\noindent
A: $\tan\varphi = \frac{\ell}{h} \leftrightarrow \ell=h\cdot\tan\varphi = 8.4\,{\rm cm}$ and $17.3\,{\rm cm}$.

\subsection{Conclusions}
Q: Why does the shortest shadow during the day point north (northern hemisphere)?\\\noindent
A: The shadow is shortest, when the Sun is highest during the day. In the northern hemisphere, this is when the Sun transits the meridian, i.e. it is exactly south. The shadow hast to point north.

\medskip\noindent
Q: What time was it, when the Sun was in the South (northern hemisphere)?\\\noindent
A: Most probably not 12 h, noon.

\medskip\noindent
Q: Why does the clock show something else than 12 h at local noon, when the Sun attains its highest elevation?\\\noindent
A: Time zones, daylight saving time, equation of time (bonus knowledge)

\medskip\noindent
Q: The length of the shadow at local noon changes throughout the year. Why?\\\noindent
A: The declination of the Sun changes. This is because the Earth with its titled axis revolves around the Sun. From the surface of the Earth, the Sun appears to be alternating between the Tropic of Cancer and the Tropic of Capricorn. This modifies the elevation of the Sun seen from any point on Earth.

\section{Conclusion}
The activity shows how the cardinal directions can be determined with the shadow of a gnomon caused by the Sun during the day. At the same time, the length of the shadow is a measure of the latitude on Earth. This knowledge represents a navigational skill that Vikings during the 8th until the 11th century used to find their course to distant destinations across the open seas. A short story is included that tells the students something about what the life of a Viking may have looked like and generates interest to study this important and influential era of European history.

\begin{acknowledgements}
This resource was developed in the framework of Space Awareness. Space Awareness is funded by the European Commission’s Horizon 2020 Programme under grant agreement no. 638653.

The hands-on activity is based on the “Parallel Earth” tool developed by Carme Alemany and Rosa Maria Ros within the educational programme EU-UNAWE.
\end{acknowledgements}

\bibliographystyle{aa}
\bibliography{Navigation}

\begin{thebibliography}{16}
\expandafter\ifx\csname natexlab\endcsname\relax\def\natexlab#1{#1}\fi

\bibitem[{Baker \& Baker(1997)}]{baker_ancient_1997}
Baker, R.~F. \& Baker, C.~F. 1997, Ancient {Greeks}: {Creating} the {Classical}
  {Tradition} (New York, USA: Oxford University Press)

\bibitem[{Berg~Petersen(2012)}]{berg_petersen_what_2012}
Berg~Petersen, I. 2012, What {Vikings} really looked like, Science Nordic,
  \url{http://sciencenordic.com/what-vikings-really-looked}

\bibitem[{Bern\'{a}th {et~al.}(2013)Bern\'{a}th, Blah\'{o}, Egri, Barta, \&
  Horv\'{a}th}]{bernath_alternative_2013}
Bern\'{a}th, B., Blah\'{o}, M., Egri, A., Barta, A., \& Horv\'{a}th, G. 2013,
  Proceedings of the Royal Society A, 469, 20130021

\bibitem[{Cunliffe(2003)}]{cunliffe_extraordinary_2003}
Cunliffe, B. 2003, The {Extraordinary} {Voyage} of {Pytheas} the {Greek},
  paperback edn. (Harmondsworth, UK: Penguin Books)

\bibitem[{Di~Cola \& Stone(2012)}]{di_cola_images_2012}
Di~Cola, J.~M. \& Stone, D. 2012, Images of {America}: {Chicago}'s1893
  {World}'s {Fair} (Charleston, South Carolina, USA: Arcadia Publishing)

\bibitem[{ESA(2013)}]{esa_esas_2013}
ESA. 2013, {ESA}'s magnetic field mission {Swarm}, European Space Agency,
  \url{http://www.esa.int/Our_Activities/Observing_the_Earth/The_Living_Planet_Programme/Earth_Explorers/Swarm/ESA_s_magnetic_field_mission_Swarm}

\bibitem[{Forte {et~al.}(2005)Forte, Oram, \& Pedersen}]{forte_viking_2005}
Forte, A., Oram, R., \& Pedersen, F. 2005, Viking {Empires} (Cambridge, UK:
  Cambridge University Press)

\bibitem[{Graham-Campbell(2001)}]{graham-campbell_viking_2001}
Graham-Campbell, J. 2001, The {Viking} {World}, 3rd edn. (London, UK: Frances
  Lincoln)

\bibitem[{Hertel(1990)}]{hertel_geheimnis_1990}
Hertel, P. 1990, Das {Geheimnis} der alten {Seefahrer}, Geographische
  {Bausteine} No.~38 (Gotha, Germany: Hermann Haack Verlagsgesellschaft mbH)

\bibitem[{Hicks(2007)}]{hicks_bayeux_2007}
Hicks, C. 2007, The {Bayeux} {Tapestry}: {The} {Life} {Story} of a
  {Masterpiece} (London, UK: Vintage Books)

\bibitem[{Johnson \& Nurminen(2009)}]{johnson_history_2009}
Johnson, D.~S. \& Nurminen, J. 2009, The {History} of {Seafaring}, 2nd edn.
  (National Geographic), authorised German Edition

\bibitem[{Nansen(1911)}]{nansen_northern_1911}
Nansen, F. 1911, In {Northern} {Mists}: {Arctic} {Exploration} in {Early}
  {Times}, Vol.~1 (New York: Frederick A. Stokes company)

\bibitem[{Sawyer(1997)}]{sawyer_oxford_1997}
Sawyer, P. 1997, The {Oxford} {Illustrated} {History} of the {Vikings}, ed.
  P.~Sawyer (Oxford, New York: Oxford University Press)

\bibitem[{Susanek(2000)}]{susanek_hygelac_2000}
Susanek, C. 2000, in Reallexikon der {Germanischen} {Altertumskunde}, 2nd edn.,
  ed. J.~Hoops, Vol.~15 (Berlin, Germany: Walter de Gruyter GmbH \& Co. KG),
  298--300

\bibitem[{Thirslund(2007)}]{thirslund_viking_2007}
Thirslund, S. 2007, Viking {Navigation} (Roskilde, Denmark: Viking Ship Museum)

\bibitem[{Tjgaard(2011)}]{tjgaard_windjamming_2011}
Tjgaard, G. 2011, Windjamming to {China} (Durham, UK: Strategic Book Group)

\end{thebibliography}

\glsaddall
\printglossaries

\begin{appendix}
\section{Supplemental material}
This unit is part of a larger educational package called “Navigation Through the Ages” that introduces several historical and modern techniques used for navigation. An overview is provided via:

\href{http://www.space-awareness.org/media/activities/attach/b3cd8f59-6503-43b3-a9e4-440bf7abf70f/Navigation\%20through\%20the\%20ages\%20compl\_z6wSkvW.pdf}{Navigation\_through\_the\_Ages.pdf}

The supplemental material is available on-line via the Space Awareness project website at \url{http://www.space-awareness.org}. The direct download links are listed as follows:
 
\begin{itemize}
  \item Worksheets: \href{https://drive.google.com/file/d/0Bzo1-KZyHftXcjV1ZUgyaHpCSU0/view?usp=sharing}{astroedu1648-Vikings-WS.pdf}
  \item Story of the Viking Galmi: \href{https://drive.google.com/file/d/0Bzo1-KZyHftXYTNFQXg0S1RHSE0/view?usp=sharing}{astroedu1648-Vikings-Story.pdf}
\end{itemize}
\end{appendix}
\end{document}